# On the Linear Stability of Magnetized Jets Without Current Sheets – Non-Relativistic Case

By

Jinho Kim[1]★, Dinshaw S. Balsara[1]★, Maxim Lyutikov[2], Sergei S. Komissarov[3], Daniel George[4] and Prasanna Kumar Siddireddy[4]

[1]*Physics Department, College of Science, University of Notre Dame, 225 Nieuwland Science Hall, Notre Dame, IN 46556, USA*

[2]*Department of Physics, Purdue University, 525 Northwestern Avenue, West Lafayette, IN 47907-2036, USA*

[3]*Department of Applied Mathematics, The University of Leeds, Leeds LS2 9GT*

[4]*Department of Physics, IIT Bombay, Powai, Mumbai 400076, India*

**Abstract**

In this paper we consider stability of magnetized jets that carry no net electric current and do not have current sheets (Gourgouliatos et al. 2012). The non-relativistic MHD equations are linearized around the background velocity and the magnetic field structure of the jet. The resulting linear equations are solved numerically inside the jet. We find that introduction of current-sheet-free magnetic field significantly improves jet stability relative to unmagnetized jets or magnetized jets with current sheets at their surface. This particularly applies to the fundamental pinch and kink modes - they become completely suppressed in a wide range of long wavelengths that are known to become most pernicious to jet stability when the evolution enters the non-linear regime. The reflection modes, both for the pinch and kink instability, also become progressively more stable with increased magnetization.

## 1. Introduction

Many astrophysical systems generate jets. The most spectacular examples are the jets from Active Galactic Nuclei (e.g. Rees 1978) and from young stars (e.g. Reipurth et al. 1998), Jets are also produced by X-ray Binaries and Gamma Ray Bursters. Although the actual mechanism of jet production is not fully established observationally, most theorists agree that it is magnetic in nature (e.g. Lovelace 1976; Blandford & Znajek 1977; Blandford & Payne 1982; Komissarov & McKinney 2007; McKinney & Narayan 2007; Komissarov & Barkov 2009; McKinney & Blandford 2009). This is partially supported by the observations of synchrotron emission from most astrophysical jets, though only very few examples of synchrotron-emitting protostellar jets are found so far. Unfortunately, these observations do not allow to measure the magnetic field strength directly, and hence to determine its dynamical importance. The total energy in magnetic field and relativistic electrons is minimized when it is equally split between these two components – this is one of the reasons why the equipartition hypothesis is so popular among astrophysicists. Additional observations providing independent information on these components, such as observations of the inverse Compton emission of the synchrotron electrons, is needed to resolve this degeneracy. Unfortunately, such information is still largely missing. The equipartition field is already sufficiently strong to influence jet dynamics. Some theoretical models of jet production and propagation predict dynamically strong magnetic field in astrophysical jets, particularly the relativistic ones.

---

★ E-mail: jkim46@nd.edu (JK); dbalsara@nd.edu (DSB)

One of the most interesting and puzzling properties of astrophysical jets is their apparent stability – they manage to keep structural integrity over huge distances. For example, jets from young stars are traced up to the distances of few parsecs. Their initial radius should be about the size of their central engine, and hence reside somewhere between the stellar radius of two solar radii and 10 AU, depending on the engine model (Ray 2012). Thus, stellar jets cover the distances of order $10^5$ or $10^7$ of their initial radius. For AGN jets the estimates are even higher, reaching $10^9$. This is in great contrast with the terrestrial and laboratory jets, which loose integrity over the distances of only $10^2$ of their initial radius. It is known that magnetic field may help to suppress some instabilities but it can also introduce new ones. These magnetic instabilities is the main reason behind the failure of many nuclear fusion projects.

The traditional way of studying instabilities of non-linear dynamical systems is via linear stability analysis. In many cases, it leads to much simpler system of linearized equations, which can be solved analytically. However, in many other cases even the linearized system does not allow general solutions in terms of analytic functions for arbitrary equilibrium configuration. One way to overcome this problem is to restrict the analysis to special equilibrium configurations which allow us to simplify the linearized system of equations even further. Accordingly, most early studies of jet stability assumed simplified jet structure, including the magnetic field topology (e.g. Hardee 1979, 1982; Cohn 1983; Payne & Cohn 1985; Istomin & Pariev 1996; Begelman 1998; Lyubarskii 1999; Tomimatsu Matsuoka & Takahashi 2001; Narayan et al. 2009). This allowed to obtain the solution in terms of Bessel and hypergeometric functions. Although very useful in many respects, this approach still cannot address the stability of jets with more complex and more realistic structure. In particular, all these equilibrium jets included surface currents, whereas Gourgouliatos et al. (2012) argued that such current sheets are likely to promote resistive jet instabilities. A jet that is free of current sheets would be free of resistive instabilities in computer simulations. As an alternative, they constructed equilibrium solutions which are current sheet-free structures. There is an intuitive reason for expecting a magnetized jet without a current sheet to be more stable – such jets carry no volumetric net current. If the surface of a magnetized jet with current sheet is perturbed, two neighboring perturbations could behave analogously to two parallel current carrying wires that are carrying current in the same direction. Since such wires attract, one might analogously expect the surface of such a jet to become more corrugated, i.e. the perturbations can grow. By avoiding current sheets on the surface, a current-sheet free jet avoids this mode of destabilization. Another interesting feature of these solutions is that the jets carry zero net current and magnetic flux. In some part of the jet cross section the poloidal electric current flows outwards and in the rest of the cross section exactly the same amount of the current flows in the opposite direction. Thus, one does not have to worry of having a return current outside of the jet on large scales. The same applies to the poloidal magnetic field. Unfortunately, the magnetic structure of these solutions is too complex for the linearized equations to allow analytical solutions.

A recent paper by Bodo et al. (2013) described a way of rectifying the deficiency of the traditional approach - when the eigenfunctions of the linear instability modes cannot be found analytically they, and the corresponding eigenvalues, have to be found numerically. In their study, they only considered jets with vanishing gas pressure ($\beta$=0) and simpler magnetic field configurations. Our approach is more general, enabling us to consider the linear stability of jets with finite gas pressure and current-sheet-free magnetic structure (Gourgouliatos et al. 2012).

The remainder of the paper is divided as follows. In Section 2 we derive the governing equations for linear stability analysis of jets with non-trivial magnetic fields and rotation. In Section 3 we describe our numerically-motivated strategy for carrying out stability analysis. In Section 4 we compare the linear stability of jets that have current sheets at their boundaries with jets that are free of current sheets at their boundaries. In Section 5 we extend our study to jets more extreme parameters, as motivated by the observations of AGN and protostellar jets. Sections 6 and 7 present discussion and conclusions.

## 2. Linearized Equations

In this paper we consider only ideal non-relativistic flows and for simplicity assume constant entropy. The last assumption, often adopted in linear stability analysis, allows us to replace the energy equation with the polytropic equation of state, i.e. $P = K\rho^{\gamma}$. Thus, the governing equations are

$$\frac{\partial \rho}{\partial t} + \nabla \cdot (\rho \mathbf{v}) = 0, \qquad (1)$$

$$\rho \frac{\partial \mathbf{v}}{\partial t} + \rho \mathbf{v} \cdot \nabla \mathbf{v} = (\nabla \times \mathbf{B}) \times \mathbf{B} - \nabla P, \tag{2}$$

$$\frac{\partial \mathbf{B}}{\partial t} = \nabla \times (\mathbf{v} \times \mathbf{B}). \tag{3}$$

To further simplify the problem, we consider only axisymmetric cylindrical non-rotating jets. In cylindrical coordinates, aligned with the jet axis, the jet solution is then described by the functions of the radial coordinate only, $\rho_0(r)$, $v_{z0}(r)$, $P_0(r)$, $B_{z0}(r)$, $B_{\phi 0}(r)$ for the mass density, axial velocity, pressure, axial and azimuthal magnetic field respectively. In fact, given the isentropy condition, the variation of mass density is completely determined by the pressure variation (see eqn. (11)). These functions are solutions of the steady-state MHD equations. In these solutions, the total (gas+magnetic) pressure in the unperturbed jet is balanced by the hoop stress of the toroidal field. In this paper, we also assume that the external gas is non-magnetized, uniform and its unperturbed velocity is zero.

Uniform jet solutions are usually parameterized by the ratio of the jet and external densities $\eta$, the jet Mach number $M$, and the magnetization parameter $\beta$, which is the ratio of the gas and magnetic pressures. For non-uniform jets these parameters become less robust as they vary across the jet. In this paper, we will be using these parameters as measured at the jet axis. For example, $\eta = \rho_j / \rho_a$, where $\rho_j$ is the jet density as measured at the jet axis and $\rho_a$ is the uniform external density in the ambient medium.

Since the steady-state solution is independent of $t$, $\varphi$ and $z$, we can Fourier expand in these coordinates and ultimately consider perturbations of the form $\delta f(t,r,\phi,z) = \delta f(r) \exp(i\omega t - im\phi - ikz)$. We perform a complex-in-time stability analysis, so that our values of "$k$" are real and our values of "$\omega$" are complex. A negative imaginary part for "$\omega$" will result in exponential growth. We make the further definition $\varpi(r) \equiv \omega - v_{z0}(r)k$ and $k_B(r) = \frac{m}{r} B_{\phi 0}(r) + k B_{z0}(r)$. The MHD equations, as well as the polytropic equation of state, and the divergence free condition ($\nabla \cdot \mathbf{B} = 0$) give us the following linearized equations:

$$i\varpi(r) \frac{\delta \rho}{\rho_0(r)} + \frac{1}{r\rho_0(r)} \frac{d}{dr}(r \rho_0(r) \delta v_r) - i\frac{m}{r} \delta v_\phi - ik \delta v_z = 0, \tag{4}$$

$$i\varpi(r) \rho_0(r) \delta v_r = -ik_B(r) \delta B_r - \frac{d(\delta \Pi)}{dr} - \frac{2}{r} B_{\phi 0}(r) \delta B_\phi, \tag{5}$$

$$i\varpi(r) \rho_0(r) \delta v_\phi = \frac{1}{r} \frac{d}{dr}(r B_{\phi 0}(r)) \delta B_r - ik B_{z0}(r) \delta B_\phi + i\frac{m}{r} B_{z0}(r) \delta B_z + i\frac{m}{r} \delta P, \tag{6}$$

$$i\varpi(r)\rho_0(r)\delta v_z + \rho_0(r)\frac{dv_{z0}(r)}{dr}\delta v_r = \frac{dB_{z0}(r)}{dr}\delta B_r + ik\, B_{\phi 0}(r)\delta B_\phi - i\frac{m}{r}B_{\phi 0}(r)\delta B_z + ik\,\delta P\,, \quad (7)$$

$$\varpi(r)\delta B_r = -k_B(r)\delta v_r\,, \tag{8}$$

$$i\varpi(r)\delta B_\phi = -\frac{d}{dr}\bigl(B_{\phi 0}(r)\delta v_r\bigr) - ik\, B_{z0}(r)\delta v_\phi + ik\, B_{\phi 0}(r)\delta v_z\,, \tag{9}$$

$$\frac{1}{r}\frac{d}{dr}(r\delta B_r) - i\frac{m}{r}\delta B_\phi - ik\,\delta B_z = 0\,, \tag{10}$$

$$\frac{\delta P}{P_0(r)} = \gamma\frac{\delta\rho}{\rho_0(r)}\,. \tag{11}$$

Here $\delta\Pi$ is the perturbation of total pressure which is defined as $\delta\Pi = \delta P + B_{\phi 0}(r)\delta B_\phi + B_{z0}(r)\delta B_z$. Note that all the unperturbed variables have subscript "0". We do not use the $B_z$-component of Eq. (3) which is redundant due to the divergence free condition. It is, therefore, replaced by the divergence-free constraint, Eq. (10).

## 3. Numerical Integration of the Linearized Equations

The linearized equations Eqs. (4)-(11) consist of four differential equations and four algebraic equations for the perturbed variables. The differential equations are Eqns. (4), (5), (9) and (10) because they contain derivatives in the perturbed variables. The algebraic equations are Eqns. (6), (7), (8) and (11) because the only derivatives that might appear in those equations are the known derivatives of unperturbed variables. However, with the help of a few manipulations that we explain in detail below, we can further reduce the number of differential equations. Physically, we anticipate that all the perturbations can be expressed in terms of the perturbation in the radial velocity, $\delta v_r$, and the perturbation in the total pressure, $\delta\Pi$. This enables us to obtain six algebraic equations. The definition of $\delta\Pi$ provides a further, seventh, algebraic equation. The result can be expressed as matrix equation:

$$\mathbf{AX} = \mathbf{B}\,, \tag{12}$$

where $\mathbf{X} = (\delta\rho, \delta P, \delta v_\phi, \delta v_z, \delta B_r, \delta B_\phi, \delta B_z)^T$, $\mathbf{A}$ is a 7×7 matrix and $\mathbf{B}$ is a column vector with 7 components that only depend on $\delta v_r$ and $\delta\Pi$. In the rest of this paragraph, we show in step-wise fashion how this is achieved.

1) Eqns. (6), (7) and (8) readily give us the first three rows of $\mathbf{AX} = \mathbf{B}$.

2) We obtain an expression for the derivative term, $d\delta v_r / dr$, from the continuity equation (eqn. 4) and substitute it in eqn. (9). This allows us to express the perturbation in the toroidal magnetic field, i.e. $\delta B_\phi$, in terms of the density and velocity perturbations. We express the resulting equation with a right hand side that depends only on $\delta v_r$. This gives us the fourth row of $\mathbf{AX} = \mathbf{B}$.

3) We differentiate eqn. (8) with respect to the radius and use it to replace the $d\delta B_r / dr$ term in eqn. (10). On further replacing the $d\delta v_r / dr$ term from the continuity equation, we obtain another equation with a right hand side that only depends on $\delta v_r$. This gives us the fifth row of $\mathbf{AX} = \mathbf{B}$.

4) Eqn. (11) gives us the sixth row and our definition of $\delta \Pi$ gives us the seventh row of $\mathbf{AX} = \mathbf{B}$.

The upshot is that eqns. (4) and (5) are two first order ordinary differential equations for the derivative of the perturbed radial velocity, $d\delta v_r / dr$, and the derivative of the perturbed total pressure, $d\delta \Pi / dr$. At any radial location within the jet, we numerically solve the system $\mathbf{AX} = \mathbf{B}$ so that all the other terms in eqns. (4) and (5) can be expressed in terms of $\delta v_r$ and $\delta \Pi$ and their radial derivatives. Consequently, given the asymptotic behavior on-axis, the perturbed variables within the jet can be numerically integrated out to all radii. (We will later show how this asymptotic behavior is obtained.)

Details of the components of matrices are provided in the Appendix A. Note that for the purposes of the matrix equation, $\mathbf{AX} = \mathbf{B}$, $\delta v_r$ and $\delta \Pi$ are input variables obtained from the two first order differential equations. All the component of $\mathbf{A}$, $\mathbf{B}$ and $\mathbf{X}$ are complex numbers, therefore we use ZGETRF and ZGETRS routine in Intel Math Kernel Library which is based on the LU decomposition.

In order to solve the differential equations for $\delta v_r$ and $\delta \Pi$, we use a Bulirsch–Stoer algorithm with adaptive step size control. To start the integration, we need to know the asymptotic behavior of the solution as $r \to 0$. There are two ways to think about this issue, one physical and one that is better rooted in mathematics. Physically, we can say that on-axis our jet has a nearly constant z-component of the magnetic field with a toroidal field that is zero. Hence the asymptotic behavior should be similar to that of a jet with a constant z-component of magnetic field. Since jets with constant z-components of magnetic field have solutions that follow the Bessel function, the jets in this paper should do the same. At a more mathematical level, in Appendix B we show that by retaining leading orders in the radius "$r$" as $r \to 0$, we can identify the leading terms in $\delta v_r$, $\delta \Pi$ and all the other flow variables. Bodo *et al.* (2013) have carried out a similar exercise for pressure-free relativistic jets when $|m| \geq 1$. We present details of this process in Appendix B because we believe our asymptotic analysis is more general. In that Appendix we show that when $|m| \geq 1$ we can take $\delta v_r \sim r^{m-1}$ and $\delta \Pi \sim C_1 r^m$ where the constant $C_1$ is also fixed by our asymptotic analysis. Similarly, when $m = 0$, we have $\delta v_r \sim r$ and $\delta \Pi \sim C_1$. These variations also match with the variation of the Bessel functions of different orders with radius. The variation of the other perturbed variables with radius is also given in Appendix B.

Bodo *et al*. (2013) integrated their equations numerically by starting with a very tiny, but non-zero, value of "*r*". Here we suggest a further improvement, drawn from the study of stellar oscillations (Cox 1980; Kim 2012). It consists of realizing that for $m \neq 0$, we rescale our variables to $\delta v_r^* = \delta v_r / r^{m-1}$, $\delta \Pi^* = \delta \Pi / r^m$. When $m = 0$, we rescale our variables to $\delta v_r^* = \delta v_r / r$, $\delta \Pi^* = \delta \Pi$. This rescaling enables us to integrate our equations by starting at $r = 0$. Furthermore, we don't have to find $d\delta v_r^* / dr$ and $d\delta \Pi^* / dr$ because they behave like even functions at $r = 0$. Realize too that if $\delta v_r^*(r = 0)$ and $\delta \Pi^*(r = 0)$ are solutions at $r = 0$ then so are $a\, \delta v_r^*(r = 0)$ and $a\, \delta \Pi^*(r = 0)$ where "*a*" is a complex number. I.e. there is an ambiguity in the interior solution up to a multiplicative constant. This ambiguity can only be resolved by applying the boundary conditions at the surface of the jet. We will describe our boundary conditions after the next paragraph.

Outside of the jet, we assume that $\rho = \rho_a$, $P = P_a$, $\mathbf{v} = 0$ and $\mathbf{B} = 0$. This condition gives one simple linearized equation for $\delta P$ which is the well-known modified Bessel equation:

$$r^2 \frac{d^2 \delta P}{dr^2} + r \frac{d\delta P}{dr} - \left( \kappa^2 r^2 + m^2 \right) \delta P = 0 \ , \tag{13}$$

where $\kappa^2 = k^2 - \frac{\rho_0}{\gamma P_0} \omega^2$. Since $\delta P$ goes to zero as $r \to \infty$, only the second kind of Bessel function is relevant, i.e. $\delta P = K_m(\kappa r)$. Note that this solution only holds when $|\arg(\kappa)| < \pi/2$ i.e. $\kappa^2$ is not real number. We use DCBKS in IMSL to get the second kind of modified Bessel function ($K_m$) with complex arguments.

At the boundary of the jet, the perturbation in the total pressure and the radial displacement from the interior and exterior have to be matched. We denote the exterior solution with a superscript of "+" and the interior solution with a superscript of "-". The matching conditions, therefore become

$$\delta \Pi^- = \delta \Pi^+ \tag{14}$$

and

$$\frac{\delta v_r^-}{i\varpi} = \frac{\delta v_r^+}{i\omega} \tag{15}$$

Recall that $\varpi(r) \equiv \omega - v_{z0}(r)k$ expresses the effect of an advected derivative. One of the above two conditions is used to resolve the fact that the interior solution is ambiguous up to a multiplicative constant. As a result, all that matters is the ratio of eqns. (14) and (15). By incorporating the modified Bessel function from the exterior solution, we get our final condition for the root solver. It is given by

$$\frac{i\varpi\delta\Pi(r=1)}{\delta v_r(r=1)} = \frac{\rho_e \omega^2 K_m(\kappa)}{\kappa K_m'(\kappa)} \tag{16}$$

Notice that when the z-component of the magnetic field in the jet is a constant, the interior solution is also represented by a modified Bessel function. In that limit, our dispersion relation in eqn. (16) reduces to eqn. (19a) of Cohn (1983, who considered the case of a uniform unmagnetized jet confined by the purely azimuthal magnetic field of its cocoon). However in the general case, the numerator and denominator on the left hand side of eqn. (16) have to be obtained via numerical integration. Because in our ODE solver we use a Bulirsch–Stoer algorithm with adaptive step control, the accuracy of solutions can be made almost as high as that dictated by the machine precision alone, which is the double precision in our calculations.

Our strategy for finding the roots of eqn. (16) is also somewhat new. Traditionally, one starts at long wavelengths where only the fundamental mode is present. As one proceeds to shorter wavelengths, the reflection modes appear and they have to be found as well. Instead, we start with the shortest wavelength, identify the roots corresponding to the fundamental and reflected modes at this wavelength, and then find the roots at longer wavelengths for each of the chosen modes separately. In order to achieve this, we first plot the absolute value of the residual of eqn. (16) as a function of ($\omega_r$, $\omega_i$) for the highest k and use this plot to locate the roots as its minima. Figure 1 presents examples of such plots for an unmagnetized uniform jet with the Mach number $M=4$ and the density ratio $\eta=0.1$ – one of the models analyzed in Cohn (1983) has the same parameters. Visual inspection of these plots allows not only to identify the fundamental and reflected modes but also to find approximate values of their roots, which are used as initial guesses for our numerical root solver of eqn. (16). Once the root corresponding to a selected mode at this shortest wavelength is found, the root solver is used to reconstruct the whole dispersion curve. During each iteration of this procedure we step towards a slightly longer wavelength and use the root value at the shorter wavelength as an initial guess. This enables us to trace out the fundamental mode as well as the reflection modes of the jet. Figure 2 shows the dispersion curves for the Cohn's model obtained in this way. Comparison with Fig. 4a from Cohn (1983) shows that we have successfully captured the *m*=0 fundamental and reflection modes. While here we considered an unmagnetized jet, for the purpose of testing only, our approach is very general and can be applied to axisymmetric jets with any magnetic field structure. In the remaining part of the paper we deal with magnetized jets which do not have current sheets. Before we proceed with presenting our results for such jets, we comment that, according to the data presented in Figure 2, the fastest growth rate of the first reflection pinch modes is significantly higher than that of the fundamental mode. For the kink modes, the fastest growth rates of the fundamental and first reflection modes are comparable. These trends continue for magnetized jets.

## 4. Stability of Current-Sheet-Free Jets

Once we have tested our numerical approach on the case with well-known analytical result, it makes perfect sense to consider a more complex flow which cannot be treated analytically. With this goal in mind, we opted to analyze the linear stability of current-sheet-free jets, whose equilibrium structure was recently studied by Gourgouliatos et al. (2012). These jets carry zero net poloidal current and the thermal gas pressure is an important dynamical component. This combination of thermal pressure and magnetic field enables us to avoid having surface currents at the jet boundary. The radial structure of the magnetic field is described by a rather complicated variant of the Grad-Shafranov equation, which in the general case can be solved only numerically. However, Gourgouliatos et al. (2012) have identified two cases when the equation becomes tractable and found two families of analytical solutions. In our work, we analyzed the linear stability of the solutions associated with use of the poloidal magnetic flux function to describe the magnetic field. The magnetic field and gas pressure of this solution are given by Equations (24)-(26) in Gourgouliatos et al. (2012). For our non-relativistic, the equations read

$$B_\phi(r) = c\alpha J_1(\alpha r) - \frac{Fr}{\alpha} , \tag{17}$$

$$B_z(r) = c\alpha J_0(\alpha r) - \frac{2F}{\alpha^2} , \tag{18}$$

$$P(r) = F\left(crJ_1(\alpha r) - \frac{Fr^2}{\alpha^2}\right) + P_0 , \tag{19}$$

where $P_0$ is gas pressure on the axis. The free parameters are set to be $c$=0.172, $\alpha$=5.14 and $F$=-1.54 which puts the jet boundary at $r$=1. For the above choice of parameters, the maximum value of $B_z$ is 1. In Fig. 3 we repeat the plots of the toroidal and axial magnetic fields from eqns. (17) and (18). These distributions can be combined with an arbitrary distribution $v_z(r)$ of the jet velocity, without upsetting the force balance. In this work we use a top hat velocity profile. The ambient pressure is constant and obtained by matching it to the pressure at the boundary of the jet. The plasma-β in the jet is, therefore, adjusted by varying the value of $P_0$.

Jets with current sheets have been studied before. The magnetic field configurations in equations (17) and (18) are certainly not unique, but they are novel. The stability of jets with this magnetic field configuration has never been studied before. The absence of a current sheet may also be very desirable for numerical simulations where numerical resistivity can masquerade as a physical resistivity. For that reason, we study it here.

**4.1 The Base Model**

In order to prominently illustrate the effect of current-sheet-free magnetic field on the jet stability, we decided to use the unmagnetized model with $M=4$ and $\eta=0.1$, whose stability we analyzed in Sec. 3 (see Figure 2) as a reference, and to retain as much of its structure as possible. In particular, we retain the constant profile for the jet velocity in all magnetic models presented here.

It is very helpful to see the results of introducing the current-sheet free magnetic field as a function of increasing magnetic field. Viewed as a progression, it becomes easier to pick out the trends. Consequently, Figures 4, 5 and 6 show the stability analysis for both the pinch ($m=0$) and kink (m=1) modes when $\beta=1$, $\beta=1/2$ and $\beta=1/4$ respectively. Figures 4, 5 and 6 are shown in the same format as in Figure 2. In all these three plots, we present the data for the fundamental mode and the first two reflection modes. Comparison of Figures 2 and 4 shows reduced growth rates in the magnetic case with $\beta=1$. The magnetic field is already dynamically important in the $\beta=1$ jet. As we increase the magnetic field strength in Figures 5 and 6, which correspond to strongly magnetized jets with $\beta=1/2$ and $\beta=1/4$, we see that the stability of the jet improves even further. The improvement in stability is particularly strong for the fundamental modes. For the fundamental pinch mode, a wide window around $kr_j=1$ appears, where this mode is not growing at all. A similar window of suppressed growth for the fundamental kink mode appears around $kr_j=4$. The results for the first two reflection modes also shows improved stability properties of the magnetic model, but now in the $kr_j \gg 1$ region, where we also observe complete suppression of these modes. Figure 6b also shows evidence for some mode mixing between the fundamental mode and the second reflection kink mode at large wave numbers, i.e. at short wavelengths. The magnetic field used in Figure 6 was strong enough to drastically alter the pressure profile of the jet, resulting in the mode mixing that we see in Figure 6b.

Figure 5 has shown that for large regions of wave number space there are no unstable modes. Our method is based on a numerical search procedure. The interested reader may well ask: How we can be sure that there are absolutely no other unstable modes in the jet? Indeed, for a numerically-motivated search process there is no ironclad way of showing that the dispersion relation has no further roots. However, it is possible to demonstrate that within a specified search space that is reasonably large, no further roots exist. (Please recall the search strategy that was described in Figure 1.) Let us focus on "k $r_j$ = 0.6" in Figure 5a. For that value of wave number, we can plot out the amplitude of our dispersion relation in a two dimensional domain given by $(\omega_r, \omega_i) \in [0,1] \times [0, 0.3]$. This is shown in the left panel of Figure 7. We see clearly that there are no growing modes. Similarly, let us focus on "k $r_j$ = 2.0" in Figure 5b. For that value of wave number, we can plot out the amplitude of our dispersion relation in a two dimensional domain given by $(\omega_r, \omega_i) \in [0,2] \times [0, 0.3]$. This is shown in the right panel of Figure 7. We can again see clearly that there is only one growing mode and that mode is the first reflection mode.

It is very interesting to ask whether the current-free aspect of the jet contributes significantly to jet stability. I.e. envision a scenario where the magnetic field configuration from eqns. (17) to (19) retained an overall helical form but the magnetic field were non-zero at the boundary of the jet. In that case, what are the changes in the jet stability? Realize, therefore, that eqns. (17) to (19) are structured so that the magnetic field goes to zero at the radius of the jet, i.e. at $r$=1. The magnetic field can be made to retain the same form but have non-zero magnetic field at the jet boundary if we replace $r := r / \xi$ with $\xi < 1$. In that case, the magnetic field will retain the same helical structure but it will have a non-zero value at the jet boundary. Fig. 8 inter-compares two magnetized jets with M=4, η=0.1 and β=1/2. The $\xi$ = 1.0 case, shown in red, is just the current-sheet free jet that we have studied before in Fig. 5 and is shown to provide a point of reference in Fig. 8. The $\xi$ = 0.7 case, shown in blue, has a current sheet at its boundary. Fig. 8a shows the stability of the fundamental and first reflection m=0, i.e. pinch, modes. Fig. 8b shows the stability of the fundamental and first reflection m=1, i.e. kink, modes. For the m=1 kink mode shown in Fig. 8b, the fundamental mode is much more stable for the current-sheet free jet, especially over a broad range of longer wavelengths. Note though that the m=0 pinch mode in Fig. 8a is slightly more stable for the jet with a current-sheet. We ascribe that to the fact that the magnetic field is parameterized by *β*, which is only evaluated at the jet's axis. As a result, the radially-averaged magnetic energy for the $\xi$ = 0.7 jet is larger than the radially-averaged magnetic energy for the $\xi$ =1 jet. Consequently, the m=0 mode in Fig. 8a has slightly greater stability for the $\xi$ =0.7 jet than for the $\xi$ =1 jet. The m=1 mode is more susceptible to instabilities driven by current sheets. As a result, the m=1 mode in Fig. 8b has substantially greater stability at long wavelengths for the $\xi$ =1 jet than the $\xi$ =0.7 jet. In all cases, we find that the first reflection mode for the $\xi$ =1 jet has improved stability compared to the $\xi$ =0.7 jet.

Figures 9 and 10 illustrates how the stability properties of the magnetic model vary with the magnetic field strength. Figure 9 shows the evolution of the fundamental pinch and kink modes. One can see that the results for low and high magnetization models are qualitatively different – whereas the dispersion curves for the low-magnetization models appears as slightly deformed versions of the non-magnetic model, the high-magnetization models show splitting of the curves into two branches separated by finite-size windows of suppressed growth. The bifurcation occurs around $\beta$=1. Outside of the windows, the growth rates show only minor changes with beta. Figure 10 shows the evolution of the first reflected pinch and kink modes. The results are reminiscent of those for the fundamental modes. Regions of suppressed growth appear in the $kr_j$ >1 zone (they may or may not continue to infinity). Outside of these windows the growth rates remain more or less unchanged.

In general, the observed instabilities can be driven either by the velocity gradient, i.e. Kelvin-Helmholtz (KH) instability, gas pressure fluctuations or the magnetic forces, i.e. Current-Driven Instability (CD) instability. For $\beta = \infty$ the jet is unmagnetized, so that we can be sure that we are dealing exclusively with the KH-instability. An increase of the growth rate for lower values of $\beta$ would be indicative of a CD-instabilities. Figures 9 and 10 show a small increase of the growth rate only at $kr_j$ ~10. Thus, one may conclude that the imposed current-sheet-free magnetic field partially suppresses the KH-instability and does not introduce strong CD-instabilities. To understand Current-Driven instabilities it is important to find the resonant surface where the resonance condition ($k \cdot B = kB_{z0} + (m/r)B_{\varphi 0} = 0$) holds. When this surface resides inside the jet, the jet becomes unstable to Current-Driven instability. Realize, therefore, that the CD-instability becomes very prominent when the magnetic field in the jet has a constant pitch angle, as is the case in the model of Bodo et al (2012). In our model, the magnetic field in the jet has a range of pitch angles, please see Fig. 3. Consequently, our model always has a resonant surface inside the jet regardless of *k* and *m* value. (The resonant surface exists even for the extremely short wavelength or *m*=-1 case.) Accordingly, both KH and CD instabilities appear in all our results. Bodo et al. (2012) have a magnetic field with a constant pitch angle. Consequently, the Current Driven instability of their model is limited to a certain wave number ($k_0 \sim 1/P$, where P is the pitch of the magnetic field). They show distinct CD instability In their figures 4-8 for highly magnetized jets. However, our model does not show dramatic difference when the CD instability becomes dominant since our model does not have a limiting wavenumber. Furthermore, please realize that the jets used in the study by Bodo et al. (2012) have zero pressure, which exaggerates the role of the CD-instability. The jets used in our study have a finite pressure which permits sound waves to carry away fluctuations. The presence of a finite sound speed, which is comparable to the Alfven speed, greatly suppresses the role of the CD-instability.

The bifurcation of dispersion curves is a particularly interesting property of our magnetic models. In order to further localize this bifurcation, we need to study the dependence on $\beta$ in greater detail. To this aim we adopted the following procedure. We pick a wavenumber in Fig. 2 and start with the real and imaginary angular frequencies of the unmagnetized jet. For that wavenumber, we progressively increase the magnetic field and re-evaluate the real and imaginary angular frequencies. This is done for increasing values of the magnetic field till the imaginary part of the frequency becomes negligibly small. Since the fundamental modes are well-separated from the first reflection modes, it is reasonable to assume that once we start with a fundamental mode, the modes that we find via this process continue to be the fundamental modes. We repeat this process for all wave-numbers that range from 0.01 to 10. As the result, we obtain the growth rate $\omega_I r_j / c_s$ as a function of two variables - $k\, r_j$ and $1/\beta$. Figure 11 shows the results for the $m=0$ and $m=1$ fundamental modes. The green dashed line in these plots shows the location where the growth rate drops below $10^{-3}$. Interestingly, for the pinch mode the bifurcation occurs almost simultaneously at two separate wave-numbers. Their corresponding windows of suppressed growth rapidly merge and form a wide single window. In contrast, for the kink mode the bifurcation occurs in a single point and the window of instability suppression remains relatively narrow.

Figure 12 shows the results for the first reflection pinch and kink modes. Unfortunately, these plots are less informative on the onset of the bifurcation - it is not clear whether it occurs at large but finite wave-number or at infinity. However, one may still conclude that at $\beta=1$ the short-wavelength reflection modes become stabilized over a wide spectral range.

## 4.2 Other models

In the previous section we studied jets with fixed parameters $M=4$, $\eta=0.1$. In part, this was dictated by the fact that one of the models studied in the seminal paper by Cohn (1983) had exactly these parameters and we could use this model as a reference point. However, AGN jets can be much lighter than the ambient gas that they propagate through, whereas protostellar jets can be even heavier than the ambient gas that they propagate through. Furthermore, the Mach numbers of both types of jets can be quite large. For that reason, the next two sub-sections explore the stability properties of jets with much larger Mach numbers than our canonical jet and with a range of density ratios.

### 4.2.1 Very Light Jet

In this section, we consider a current-sheet free jet with $M=10$, $\eta=0.01$. These parameters are closer to those of AGN jets compared to the base model. We repeated all the steps that we undertook in Sub-section 4.1 except that in this Sub-section we consider the current-sheet free jet with $M=10$, $\eta=0.01$. The results in this Sub-section are presented in the same format as the previous Sub-section. Figures 13, 14, 15 and 16 are to be compared with Figures 9, 10, 11 and 12 of the Section 4.2 respectively. Inspection of the data shows that the two major trends found for the base model are repeated for the very light jet:- 1) The growth rates of unstable modes are normally reduced with increased magnetization. (At some wavenumbers the growth rate of reflection modes may actually increase but only weakly.) and 2) When the magnetic field becomes dynamically important, i.e. at around $\beta=1$, windows of fully suppressed instability appear.

The most significant quantitative differences with the base model are observed in the data for fundamental modes. One can see that already in the non-magnetic case, the growth rates are systematically lower. For magnetic models, the windows of suppressed instability are substantially broader – for $\beta$ =0.5 the instability is suppressed at a range of $0.051 < kr_j < 8.4$ (m=0), $0.40 < kr_j < 9.1$ (m=1).

For reflection modes, the growth rates are less affected. The domain of their instability is shifted towards lower wavenumbers. The bulk of this shift is already present in the non-magnetic case and hence can be attributed to the properties of KH-instability.

### 4.2.2 Heavy Jet

In our last example we consider heavy jet with *M*=20, $\eta$=10. Just as we did for the very light jet, we repeated all the steps of base model study and presented the results in the same format (see Figures 17-20). Inspection of these plots shows the same trends again. The magnetic field keeps playing the same role as before in reducing the growth rate of unstable modes and creating windows of suppressed instability. The bifurcations occur at $\beta$~1 again. In contrast to the very light jets, these windows are now somewhat narrower than in the base model. However, the domain of instability for first reflection modes is still shifted towards lower wavenumbers.

### 5. Discussion

Stability is undoubtedly one of the key issues in the physics of cosmic jets, which has many sides to it. Instabilities are likely to lead to dissipation of both the bulk kinetic and magnetic energies. The dissipated energy can then be emitted via different thermal and non-thermal mechanism and hence the stability issue is tightly connected to the problem of observed emission. Instabilities can result in jets developing large and small scale structures like wiggles, helical patterns, knots etc. and thus relates to the issue of the observed jet morphology. But probably the most important of all is the issue of jet survival. As we have already discussed in the Introduction, in contrast to their terrestrial and laboratory counterparts, the astrophysical jets exhibit remarkable ability to preserve their integrity over huge distances – distances that can exceed the initial jet radius by up to a billion times! In extragalactic radio sources of type 2 in the Fanaroff-Riley classification, AGN jets can be traced all the way up to the leading hot spots, which are hundreds of kiloparsecs away from galactic nuclei. In the type 1 sources, the AGN jets are shorter and are seen to turn into what appears to be slow turbulent plumes. This transition is reminiscent of what happens to terrestrial supersonic jets due to KH-instability.

Linearization of governing equations is the traditional way of studying stability of dynamical systems. The problem is then reduced to the eigenvalue problem for linear equations, which are significantly simpler compared to the original nonlinear ones. This is a very powerful mathematical tool, particularly when the eigenfunctions can be expressed in terms of well-known analytic functions. Unfortunately, this is not always the case and to achieve solutions to the linearized system one is often forced to consider rather over-simplified equilibrium configurations. In more general setting, the eigenvalue problem has to be solved numerically. This is the way the linear stability analysis of astrophysical jets is currently evolving. Following Bodo et al. (2013), and also significantly amplifying on that work, we have applied this approach to cylindrical magnetized jets free of surface currents (Gourgouliatos et al. 2012) and demonstrated its robustness and efficiency. We have found that such magnetic field inhibits growth of KH-instability modes and does not introduce strong CD-instabilities. When the magnetic field exceeds the equipartition strength, the instabilities become fully suppressed for a whole range of wavelengths. This is particularly significant for the so-called fundamental modes, which in some of our cases become suppressed for wavelengths ranging from a fraction to a hundred of jet radii. However, not all modes are suppressed and strictly-speaking the jets remain linearly unstable.

The accepted wisdom says that unstable equilibrium solutions cannot describe natural phenomena – they are self-destructing. However they can still be relevant as approximations of reality. For example, perturbations, which grow rapidly while their amplitude is small, may saturate in the nonlinear regime (e.g. O'Neill, Beckwith & Begelman 2012). Moreover, unstable solutions may remain near the equilibrium for some time even if eventually they move far away from it. Only numerical solution of original nonlinear equations can answer these question. This is why numerical simulations have become a popular tool for stability studies (see Mizuno et al. 2012; O'Neill et al. 2012; Poth & Komissarov 2014 for some of the recent examples). This is one of the ways our study of current-sheet-free jets will have to be continued eventually.

As far as the linear analysis is concerned, one of the main limitations of the present study is the neglect of the velocity shear within the jet. We expect that a sheared jet will show even better stability properties, as shear tends to destroy coherence of perturbations and suppresses small scale instabilities (Michalke 1964, Chandrasekhar 1961, Blumen, Drazin & Billings 1975). The study of shear in the jet is deferred to a subsequent paper. This study also needs to be extended to the relativistic regime, and we defer that also to a subsequent paper.

In this paper we have addressed the long-term linear stability of magnetized jets propagating through a constant density medium. We expect that these calculations are applicable to the largest scales of astrophysical jets from Active Galactic Nucleii, i.e. on the scales of tens to hundreds of kiloparsec, like those in the famous radio galaxy Pictor A. We have studied the stability of such a possible cylindrical magnetized jet configuration.

As they propagate through realistic environments, cosmic jets are not quite cylindrical but exhibit a certain amount of lateral expansion. In the super-fast-magnetosonic regime, the speed of steady-state non-relativistic jets remains almost constant and due to the magnetic flux conservation the azimuthal component of magnetic field decreases as $1/r_j$, where poloidal component as $1/r_j^2$, which is much faster. Since the radius of astrophysical jets increases by many orders of magnitude, especially at the base of the jet where it is launched, it is inconceivable that the poloidal component components is the same order as the azimuthal one everywhere along the jet. For magnetically-accelerated jets, the azimuthal component becomes of the order as the poloidal one at the Alfven surface, which is only a little bit closer to the origin than the fast-magnetosonic surface. The dominance of the toroidal magnetic field will trigger current-driven instabilities with the jet, like sausage and the kink modes. As a result, some toroidal magnetic field will be destroyed. We hypothesize that after entering a nearly-constant density profile in the intergalactic medium a narrow AGN jet finds a cylindrical equilibrium configuration with similar toroidal and poloidal magnetic fluxes. This provides a natural scenario where extragalactic jets could naturally relax towards the magnetic field configurations that we have explored here.

For relativistic jets, the transition from jet launching to free propagation can be even more interesting. In this regime one has to distinguish between the magnetic field as measured in the source frame and in the jet frame. For the jet magnetostatic equilibrium it is the jet frame magnetic field which matters. While the poloidal component of this field still varies as $1/r_j^2$, the azimuthal one varies as $1/(r_j \gamma_j)$, where $\gamma_j$ is the jet Lorentz factor as measured in the source frame. Studies of relativistic magnetized jets show that they continue to accelerate in the super-fast-magnetosonic regime – their Lorentz factor increases (Komissarov et al., 2009). When such jets are confined by an external gas with total pressure $P_{tot} \sim z^{-\kappa}$, where $z$ is the distance along the jet and 0<$\kappa$<2, the Lorentz factor grows as $\sim r_j$ (Komissarov et al., 2009) and thus the azimuthal magnetic field decreases at the same rate as the poloidal one. Because of this effect, an equilibrium configuration with comparable poloidal and azimuthal components of magnetic field can be maintained for longer.

One additional interesting possibility for enhancing the stability of extragalactic jets is that the lateral expansion of astrophysical jets helps to stabilize them. Indeed, such an expansion constantly increases the communication time across the jet and thus slows down the development of instabilities. When $\kappa$>2 the causal communication across the jets is completely lost and the jets should become absolutely stable to global instabilities. Recent numerical simulations provide very nice support for this idea (Porth & Komissarov 2014).

## 6. Conclusions

Normally, linear stability analysis of astrophysical jets is based on simplifying assumptions about the jet's velocity profile and magnetic field structure - the jet's velocity is usually taken to be a top-hat velocity profile and its magnetic field either constant in the longitudinal direction or restricted to loops of magnetic field. The simplifications are needed in to obtain closed-form analytical representation for the perturbed variables in the jet. These analytical functions usually take the form of modified Bessel functions or hypergeometric functions. In this paper we adopt a different, numerically-motivated approach to linear stability. With this approach, the jet is allowed to have any velocity profile and any unperturbed magnetic field structure.

Specifically, we focus on magnetic field structures that are free of current sheets on the surface of the jet (Gourgouliatos et al. 2012). We believe that these magnetic field structures are more realistic and confer some better stability properties to the jet. The non-relativistic magnetohydrodynamic (MHD) equations are linearized around a general velocity profile in the jet and a general magnetic field in the jet. The resulting linear equations are solved numerically inside the jet. At the radial boundary of the jet, we follow convention and match the pressure and displacement from inside the jet to the corresponding analytical solution in the ambient medium.

We find that current-sheet-free magnetic field can significantly reduce the jet instability compared to the non-magnetic case. For weak magnetic fields (defined in terms of plasma-beta, i.e. the ratio of gas to magnetic pressure) the jets display the well-known fundamental and reflection modes for the pinch and kink instabilities. However, as the magnetic field is increased slightly past equipartition, the stability properties of these modes improve. The most dramatic improvement, both for pinch and kink instabilities, occurs in the fundamental modes, particularly at long wavelengths. These are the very wavelengths that are known to become most pernicious to jet stability when the evolution enters the non-linear regime. The reflection modes, both for the pinch and kink instability, also become progressively more stable with decreasing plasma-beta, but to a lesser degree.

**Acknowledgements**

DSB acknowledges support via NSF grants NSF-AST-1009091 , NSF-ACI-1307369 and NSF-DMS-1361197. DSB and ML also acknowledge support via NASA grants from the Fermi program as well as NASA-NNX 12A088G. Several simulations were performed on a cluster at UND that is run by the Center for Research Computing. Computer support on NSF's XSEDE and Blue Waters computing resources is also acknowledged.

**Appendix A: Components of the matrix A and B**

All the components of the matrix $\mathbf{A}$ and column vector $\mathbf{B}$ for solving seven algebraic equations ($\mathbf{AX} = \mathbf{B}$, Eq. 12) described in section III are given by

$$\mathbf{A} = \begin{bmatrix} 0 & -i\frac{m}{r} & i\varpi\rho_0 & 0 & -\left(\frac{dB_{\phi 0}}{dr} + \frac{B_{\phi 0}}{r}\right) & ikB_{z0} & -i\frac{m}{r}B_{z0} \\ 0 & -ik & 0 & i\varpi\rho_0 & -\frac{dB_{z0}}{dr} & -ikB_{\phi 0} & i\frac{m}{r}B_{\phi 0} \\ 0 & 0 & 0 & 0 & \varpi & 0 & 0 \\ -i\varpi\frac{B_{\phi 0}}{\rho_0} & 0 & ik_B & 0 & 0 & i\varpi & 0 \\ -i\frac{k_B}{\rho_0} & 0 & -i\frac{m}{r}\frac{k_B}{\varpi} & -ik\frac{k_B}{\varpi} & \frac{1}{r}+\frac{k}{\varpi}\frac{dv_{z0}}{dr} & -i\frac{m}{r} & -ik \\ -\gamma P_0 & \rho_0 & 0 & 0 & 0 & 0 & 0 \\ 0 & -1 & 0 & 0 & 0 & -B_{\phi 0} & -B_{z0} \end{bmatrix},$$

$$\mathbf{B} = \begin{bmatrix} 0 \\ -\rho_0 \dfrac{dv_{z0}}{dr}\delta v_r \\ -k_B \delta v_r \\ \left( \dfrac{B_{\phi 0}}{\rho_0}\dfrac{d\rho_0}{dr} + \dfrac{B_{\phi 0}}{r} - \dfrac{dB_{\phi 0}}{dr} \right)\delta v_r \\ \dfrac{k_B}{\varpi}\left( \dfrac{C}{k_B} - \dfrac{D}{rk_B} - \dfrac{1}{\rho_0}\dfrac{d\rho_0}{dr} \right)\delta v_r \\ 0 \\ -\delta\Pi \end{bmatrix}, \quad \mathbf{X} = \begin{bmatrix} \delta\rho \\ \delta P \\ \delta v_\phi \\ \delta v_z \\ \delta B_r \\ \delta B_\phi \\ \delta B_z \end{bmatrix} \quad (\text{A1})$$

where $C = \dfrac{m}{r}\dfrac{dB_{\phi 0}}{dr} + k\dfrac{dB_{z0}}{dr}$ and $D = kB_{z0} + \dfrac{2m}{r}B_{\phi 0}$. Note that all components does not contain derivatives of perturbation variables and $\delta v_r$ and $\delta\Pi$ only appear in $\mathbf{B}$

**Appendix B: Asymptotic behavior of solutions at small radii.**

i) $m = 0$

The power lows of all the perturbation variables deduced from linearized equation by the assumption of $\delta\rho \sim r^\alpha$ near $r = 0$ are provided as:

$\delta\rho = \delta\rho^* r^\alpha$; $\delta P = \delta P^* r^\alpha$; $\delta v_r = \delta v_r^* r^{\alpha+1}$; $\delta v_\phi = \delta v_\phi^* r^{\alpha+1}$; $\delta v_z = \delta v_z^* r^\alpha$;

$\delta B_r = \delta B_r^* r^{\alpha+1}$; $\delta B_\phi = \delta B_\phi^* r^{\alpha+1}$; $\delta B_z = \delta B_z^* r^\alpha$.

Up to leading order (after cancelling out leading order of $r$) Eqs. (4) - (11) become

$$i\varpi\dfrac{\delta\rho^*}{\rho_0} + (\alpha+2)\delta v_r^* - ik\delta v_z^* = 0, \quad (\text{B1})$$

$$\alpha(B_{z0}\delta B_z^* + \delta P^*) = 0, \quad (\text{B2})$$

$$i\varpi\rho_0\delta v_\phi^* = 2B_{\phi 0}'\delta B_r^* - ikB_{z0}\delta B_\phi^*, \quad (\text{B3})$$

$$i\varpi\rho_0\delta v_z^* = ik\delta P^*, \quad (\text{B4})$$

$$\varpi\delta B_r^* = -k_B\delta v_r^*, \quad (\text{B5})$$

$$i\varpi\delta B_\phi = -(\alpha+2)B_{\phi 0}'\delta v_r^* - ikB_{z0}\delta v_\phi^* + ikB_{\phi 0}'\delta v_z^*, \quad (\text{B6})$$

$$(\alpha+2)\delta B_r^* = ik\delta B_z^*, \tag{B7}$$

$$\frac{\delta P^*}{P_0} = \gamma \frac{\delta \rho^*}{\rho_0}, \tag{B8}$$

Since $\delta B_z^*$ and $\delta P^*$ are expressed in terms of $\delta v_r^*$, $\alpha$ must be 0 for the non-trivial solution of $\delta v_r^*$. By substituting $\delta \rho^*$ and $\delta v_z^*$ in Eq. (1) for $\delta P^*$ using Eqs. (B4) and (B8), we can obtain an expression of $\delta P^*$ in terms of $\delta v_r^*$, i.e.

$$\delta P^* = \frac{2i\delta v_r^*}{\dfrac{\varpi}{\gamma P_0} - \dfrac{k^2}{\varpi \rho_0}}. \tag{B9}$$

In addition, Eqs. (B5) and (B7) give

$$\delta B_z^* = \frac{2iB_{z0}}{\varpi}\delta v_r^*. \tag{B10}$$

Since the definition of total pressure perturbation is $\delta\Pi^* = \delta P^* + B_{\phi 0}'\delta B_\phi^* r^2 + B_{z0}\delta B_z^*$, we can obtain $\delta\Pi^*(r=0) = B_{z0}\delta B_z^* + \delta P^*$ in terms of $\delta v_r^*$ by making substitutions of Eqs. (B9) and (B10).

ii) $|m|\geq 1$

Like m=0, the perturbed variables have following relations near $r=0$:

$\delta\rho = \delta\rho^* r^\alpha$; $\delta P = \delta P^* r^\alpha$; $\delta v_r = \delta v_r^* r^{\alpha-1}$; $\delta v_\phi = \delta v_\phi^* r^{\alpha-1}$; $\delta v_z = \delta v_z^* r^\alpha$;

$\delta B_r = \delta B_r^* r^{\alpha-1}$; $\delta B_\phi = \delta B_\phi^* r^{\alpha-1}$; $\delta B_z = \delta B_z^* r^\alpha$.

After substitution of above relation in Eqs. (4) - (11) provides

$$\alpha\delta v_r^* - im\delta v_\phi^* = 0, \tag{B11}$$

$$i\varpi\rho_0\delta v_r^* = -ik_B^*\delta B_r^* - \alpha\delta\Pi^* - 2B_{\phi 0}'\delta B_\phi^*, \tag{B12}$$

$$i\varpi\rho_0\delta v_\phi^* = 2B_{\phi 0}'\delta B_r^* - ikB_{z0}\delta B_\phi^* + im(B_{z0}\delta B_z^* + \delta P^*), \tag{B13}$$

$$i\varpi\rho_0\delta v_z^* + \rho_0 v_{z0}''\delta v_r^* = B_{z0}''\delta B_r^* + i(k-m)B_{\phi 0}'\delta B_\phi^* + ik\delta P^*, \tag{B14}$$

$$\varpi \delta B_r^* = -k_B^* \delta v_r^*, \tag{B15}$$

$$i\varpi \delta B_\phi^* = -\alpha B_{\phi 0}' \delta v_r^* - ikB_{z0} \delta v_\phi^*, \tag{B16}$$

$$\alpha \delta B_r^* - im\delta B_\phi^* = 0, \tag{B17}$$

$$\frac{\delta P^*}{P_0} = \gamma \frac{\delta \rho^*}{\rho_0}, \tag{B18}$$

where $k_B^* = mB_{\phi 0}' + kB_{z0}$ and $\delta \Pi^* = \delta P^* + B_{\phi 0}' \delta B_\phi^* + B_{z0} \delta B_z^*$. Eqs. (B11), (B15), (B16) and (B17) give the expressions of $\delta v_\phi^*$, $\delta B_r^*$ and $\delta B_\phi^*$ in terms of $\delta v_r^*$. Furthermore, making substitutions of $\delta B_r^*$ and $\delta B_\phi^*$ in $\delta \Pi^*$ definition also gives $\delta \Pi^*$ dependent only on $\delta v_r^*$. We make a further substitution of last term of angular momentum equation (Eq. B13) to $\delta \Pi^* - B_{\phi 0}' \delta B_\phi^*$ and finally obtain an expression in terms of $\delta v_r^*$:

$$\frac{(k_B^2 - \rho_0 \varpi^2)(m^2 - \alpha^2)}{\alpha m \varpi} \delta v_r^* = 0. \tag{B19}$$

For the non-trivial solution of $\delta v_r^*$, $\alpha = m$ or $-m$. Since all the perturbations are regular at $r = 0$, we only can take a solution of $\alpha = |m|$.

By substituting $\delta B_r^*$ and $\delta B_\phi^*$ in Eq. (B12) for $\delta v_r^*$ using Eqs. (B11), (B15) and (B16), we can obtain an expression of $\delta \Pi^*$ in terms of $\delta v_r^*$, i.e.

$$\delta \Pi^* = \frac{i}{|m|\varpi}\left[ \left(k_B^*\right)^2 - \varpi^2 \rho_0 - 2\frac{|m|}{m} k_B^* B_{\phi 0}' \right] \delta v_r^* \tag{B21}$$

This completes our discussion of the asymptotic behavior of the jet at small radii.

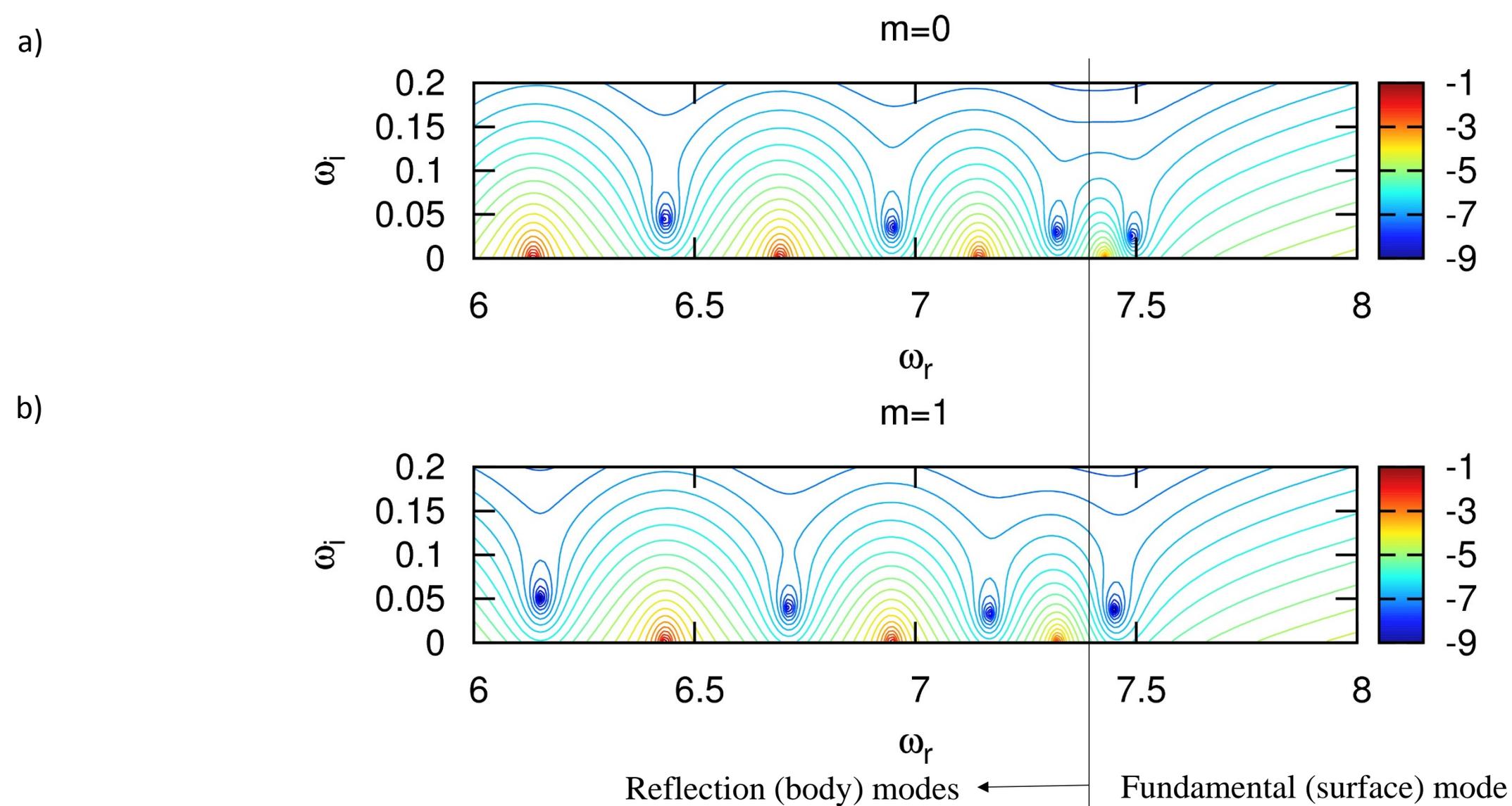

Fig. 1 shows the search space that is used for finding fundamental and reflection modes. The unmagnetized jet has $M=4$ and $\eta=0.1$. A range of values in ($\omega_r$, $\omega_i$) is selected and the amplitude of the complex dispersion relation is plotted for that range. The roots of the dispersion relation are easily identified as the locations where the amplitude vanishes. The axes show the complex frequency plane for $m=0$ (top) and $m=1$ (bottom). Fundamental (surface) mode and the first three reflection (body) modes are found via this search strategy.

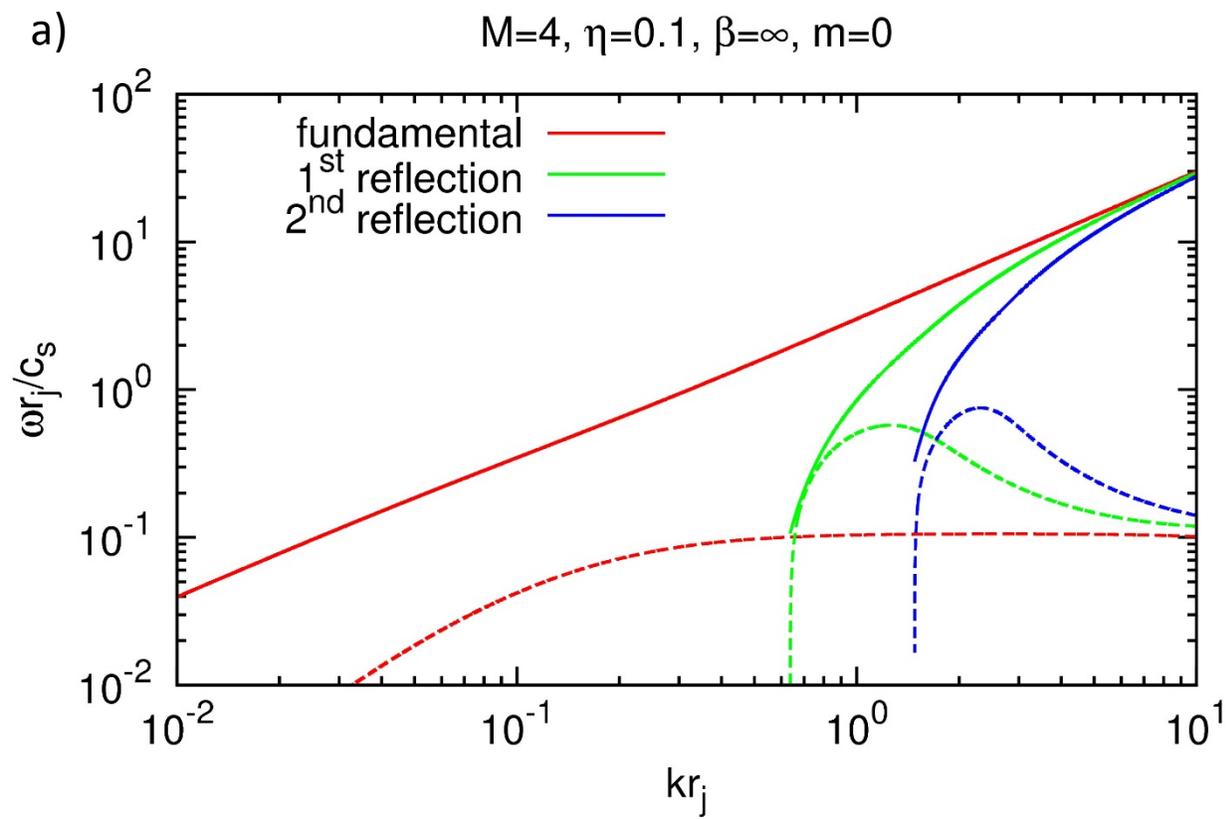 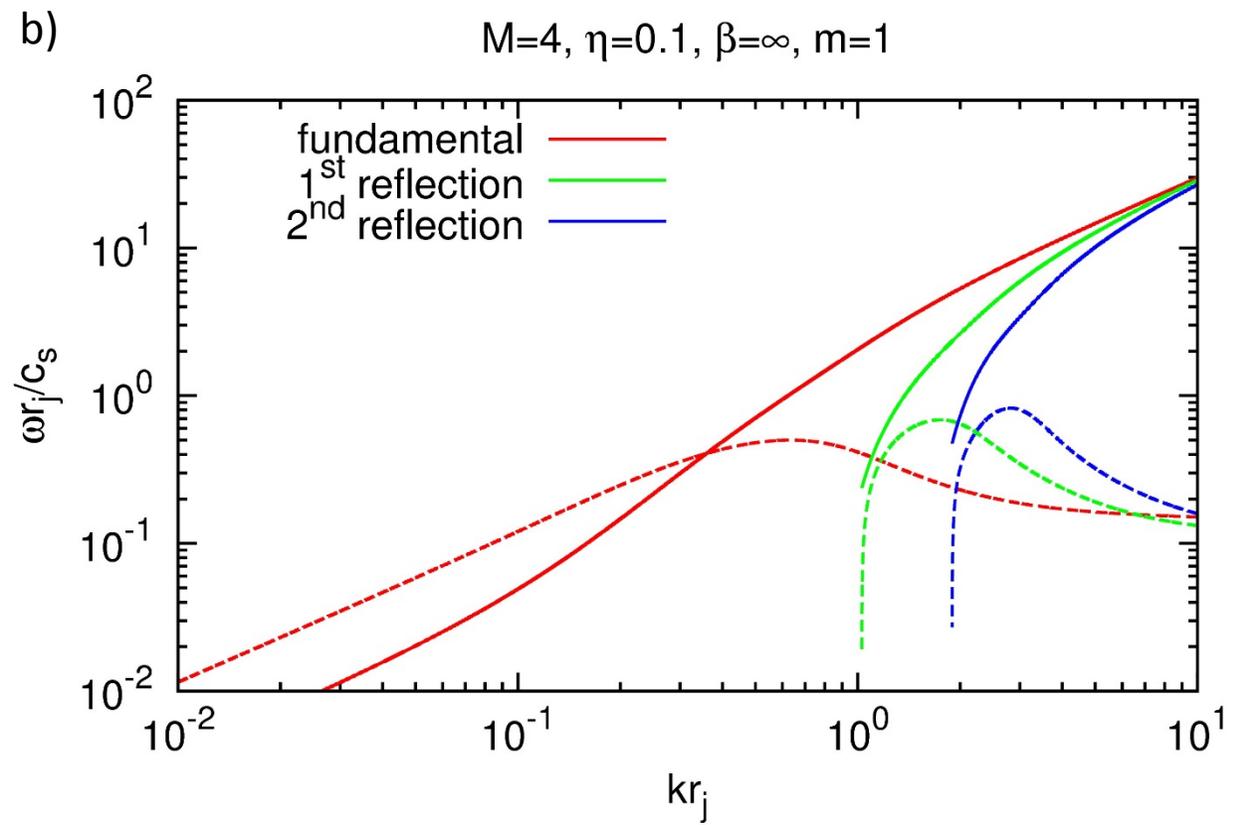

*Fig. 2. Angular frequency (solid line) and temporal growth rate (dashed line) versus longitudinal wavenumber k for pinching (m=0, left) and helical (m=1, right) modes of non-magnetized jet. The unmagnetized jet has M=4 and η=0.1. Fig. 2a should be compared to Fig. 4a from Cohn (1983).*

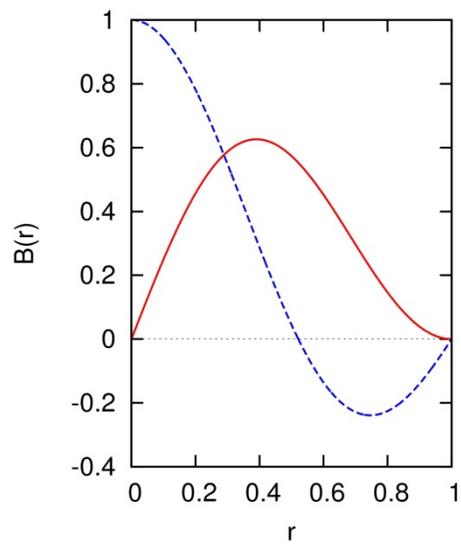

*Fig 3 from Gourgouliatos et al (2012) shows the toroidal magnetic field (red solid line) and the axial field (blue dashed line) as a function of the jet radius. Notice that the fields are zero at the jet boundary, resulting in jets that do not have a current sheet at the boundary.*

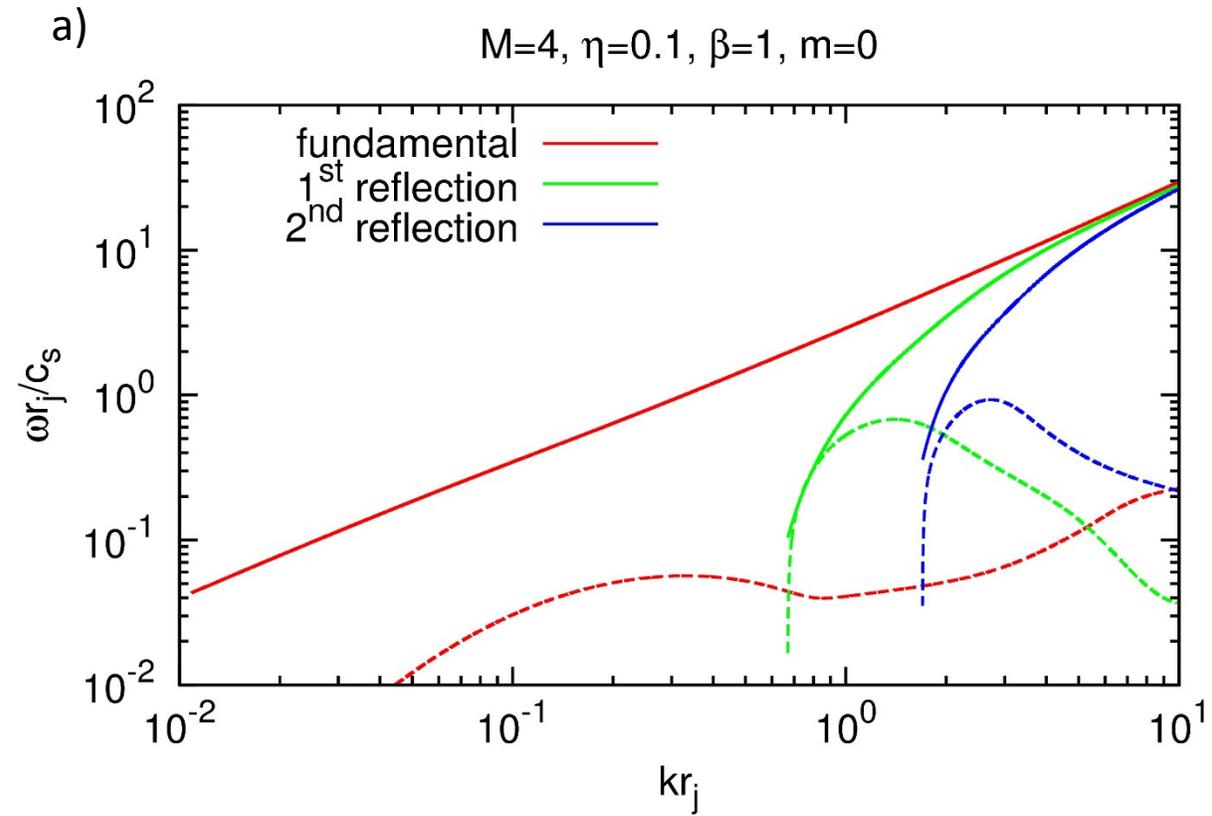 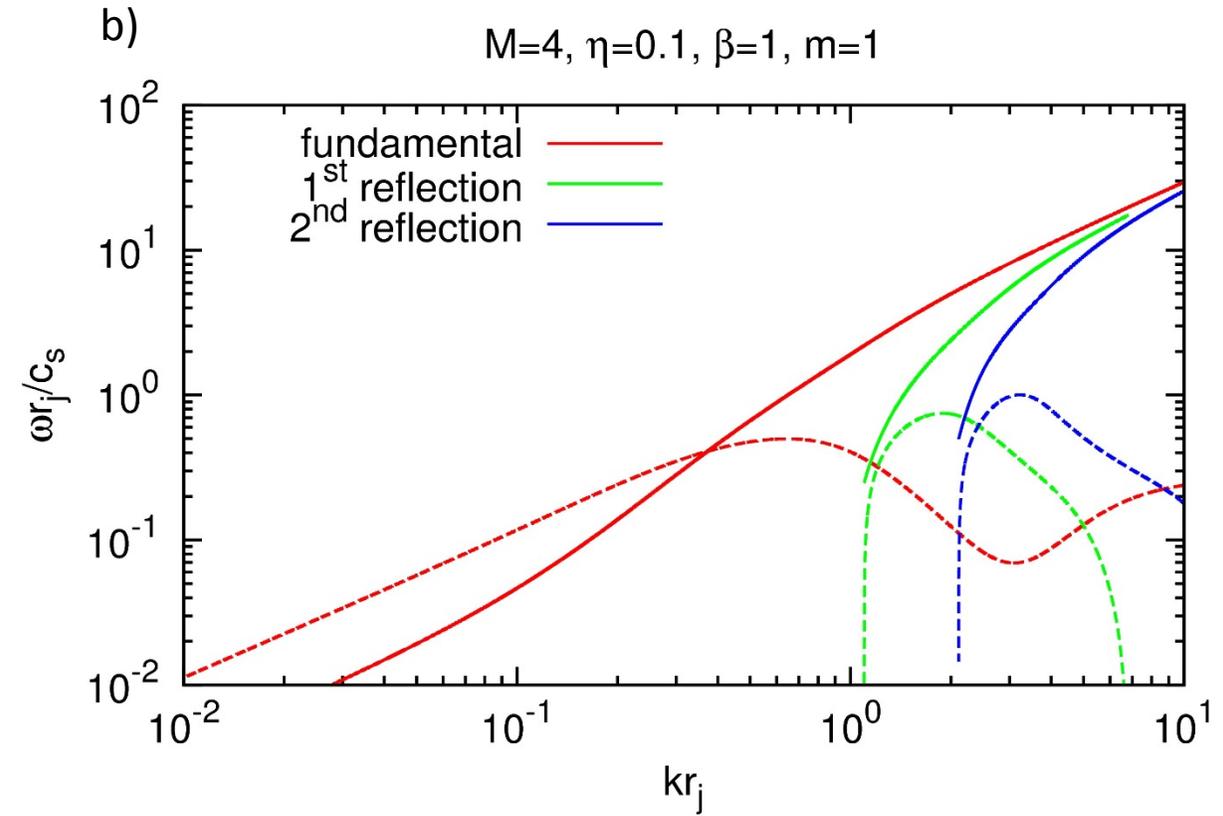

*Fig. 4 corresponds to a magnetized current-sheet free jet with M=4, η=0.1 and β=1; i.e. the on-axis magnetic pressure is in equipartition with the gas pressure. Fig 4a shows the angular frequency (solid line) and temporal growth rate (dashed line) for the m=0 mode while Fig. 4b shows the same for the m=1 mode. The fundamental mode and first two reflection modes are shown.*

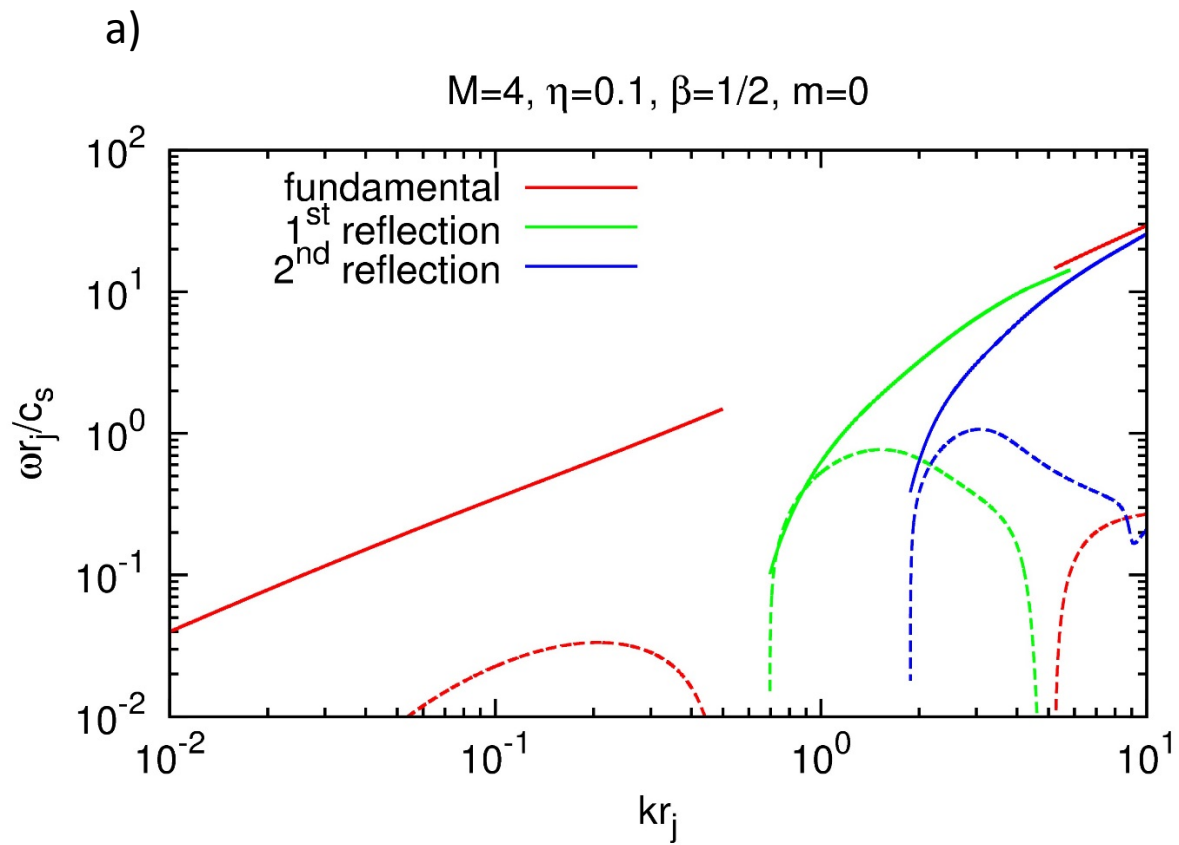 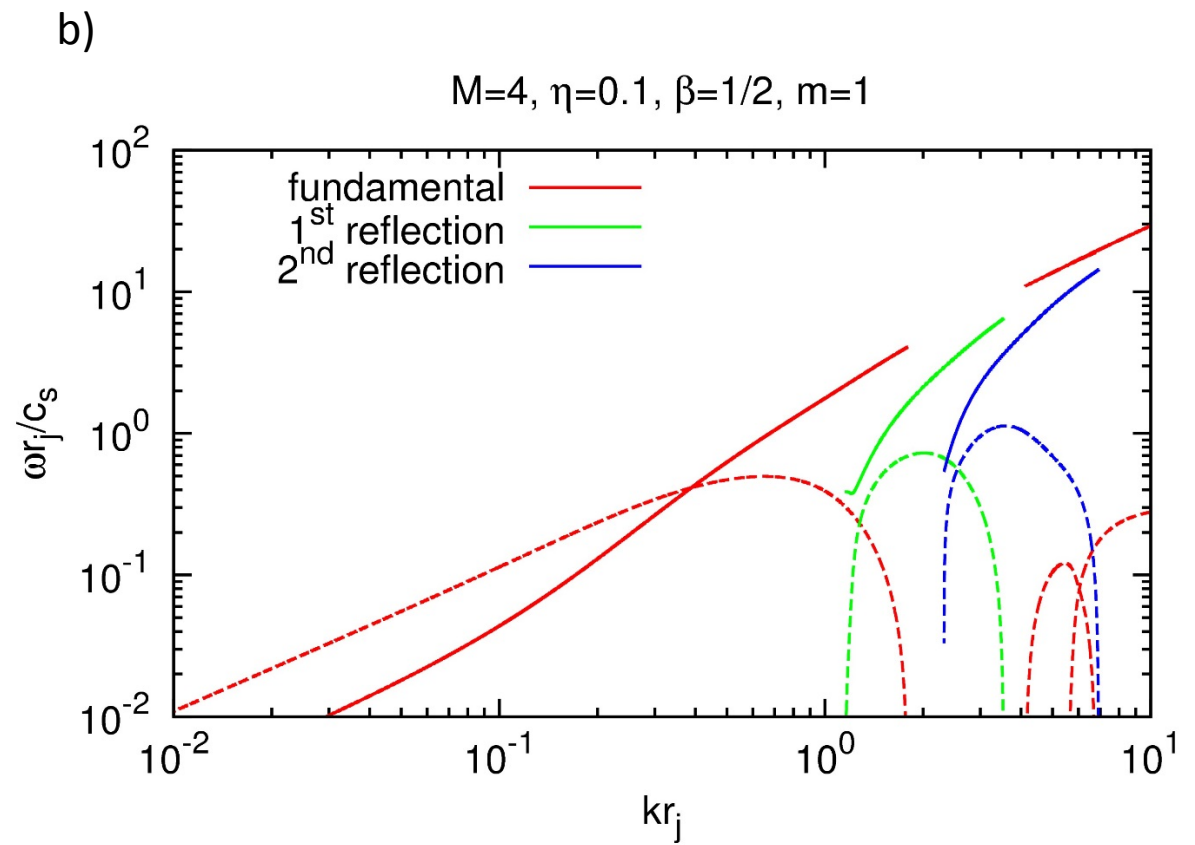

*Fig. 5 corresponds to a magnetized current-sheet free jet with $M=4$, $\eta=0.1$ and $\beta=1/2$; i.e. the on-axis magnetic pressure is twice the gas pressure. Fig 5a shows the angular frequency (solid line) and temporal growth rate (dashed line) for the $m=0$ mode while Fig. 5b shows the same for the $m=1$ mode. The fundamental mode and first two reflection modes are shown.*

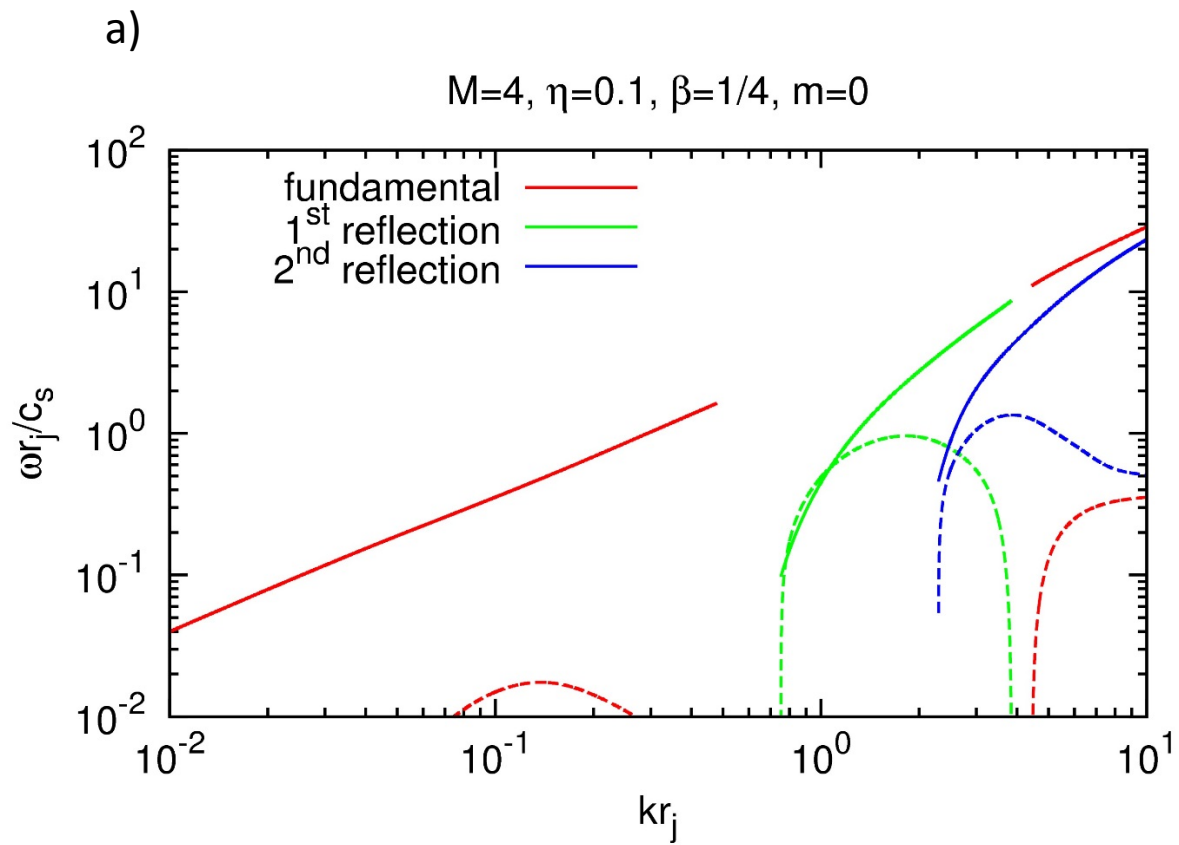 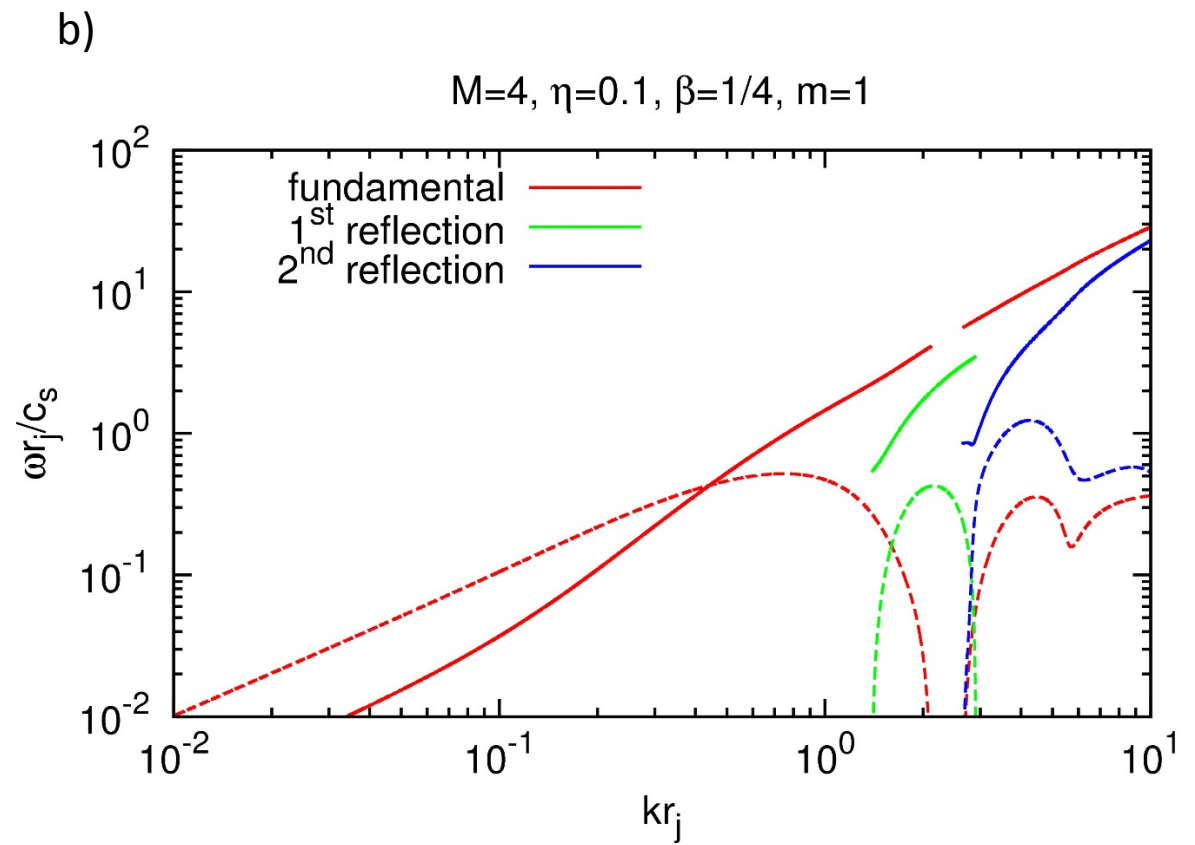

Fig. 6 corresponds to a magnetized current-sheet free jet with $M=4$, $\eta=0.1$ and $\beta=1/4$; i.e. the on-axis magnetic pressure is four times the gas pressure. Fig 6a shows the angular frequency (solid line) and temporal growth rate (dashed line) for the $m=0$ mode while Fig. 6b shows the same for the $m=1$ mode. The fundamental mode and first two reflection modes are shown.

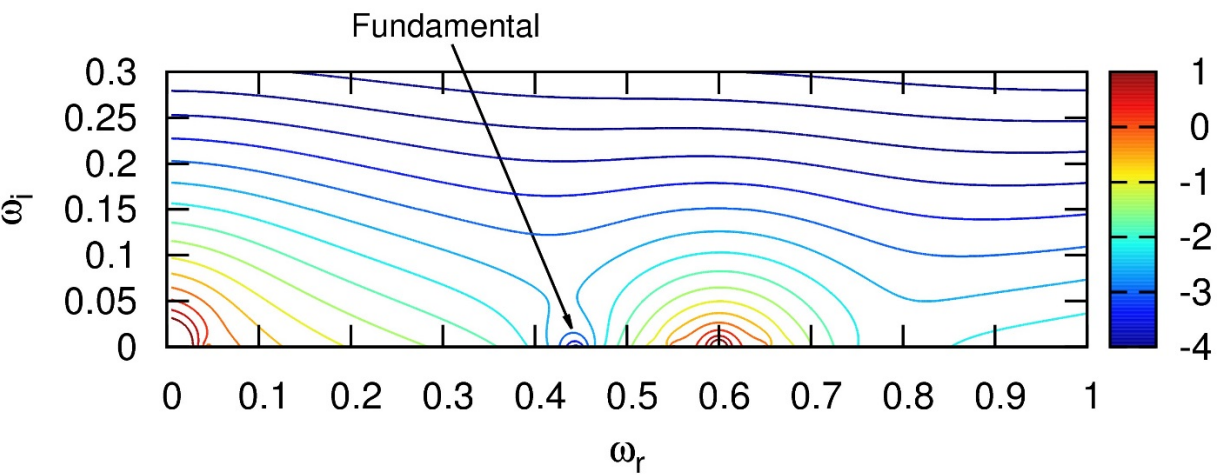 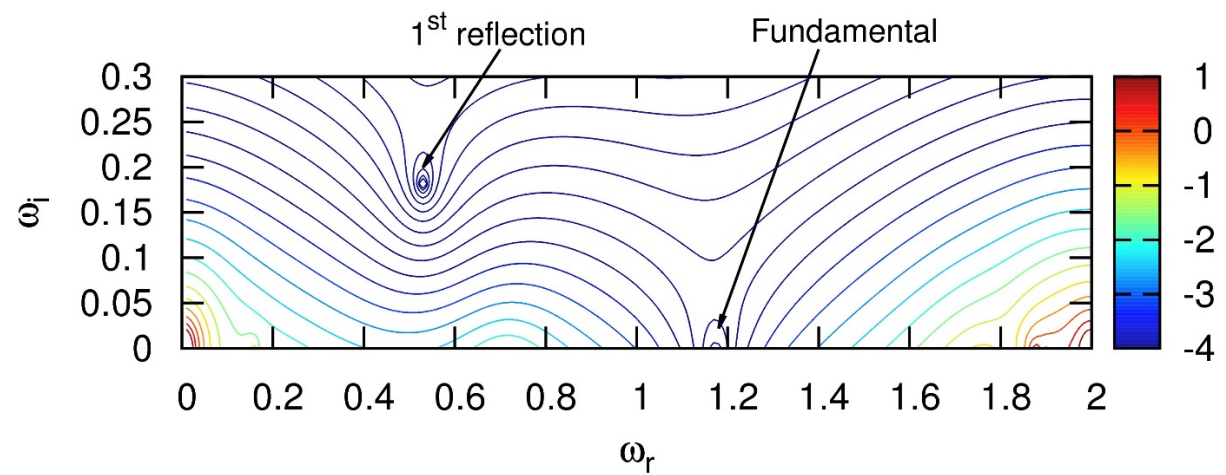

*Fig. 7, which is analogous to Fig. 1, shows the search space that is used for finding fundamental and reflection modes in Fig. 5. A range of values in ($\omega_r$, $\omega_i$) is selected and the amplitude of the complex dispersion relation is plotted for that range. The roots of the dispersion relation are easily identified as the locations where the amplitude vanishes. Fig. 7a corresponds to "k $r_j$ = 0.6" in Fig. 5a. By scanning the colors of the contours we see, therefore, that there is only one fundamental pinch mode. Fig. 7b corresponds to "k $r_j$ = 2.0" in Fig. 5b. By scanning the colors of the contours we see, therefore, that there is one fundamental and one reflection kink mode.*

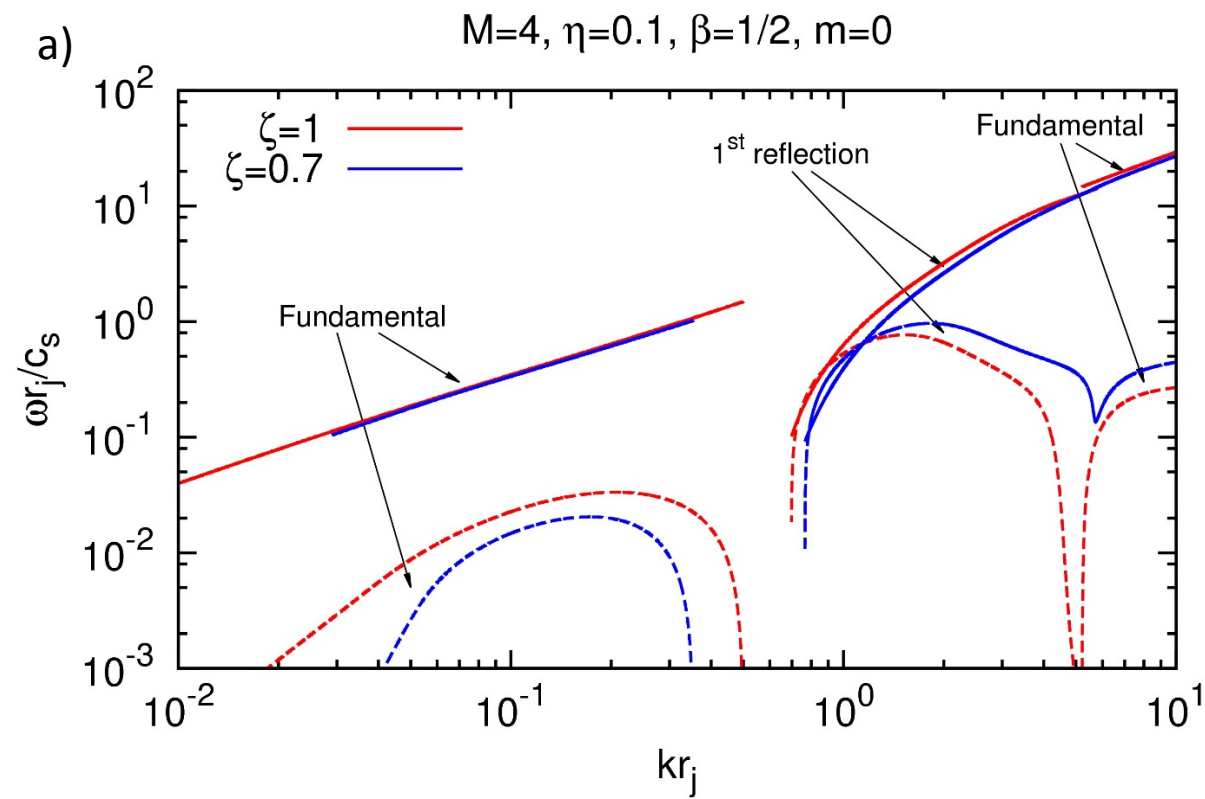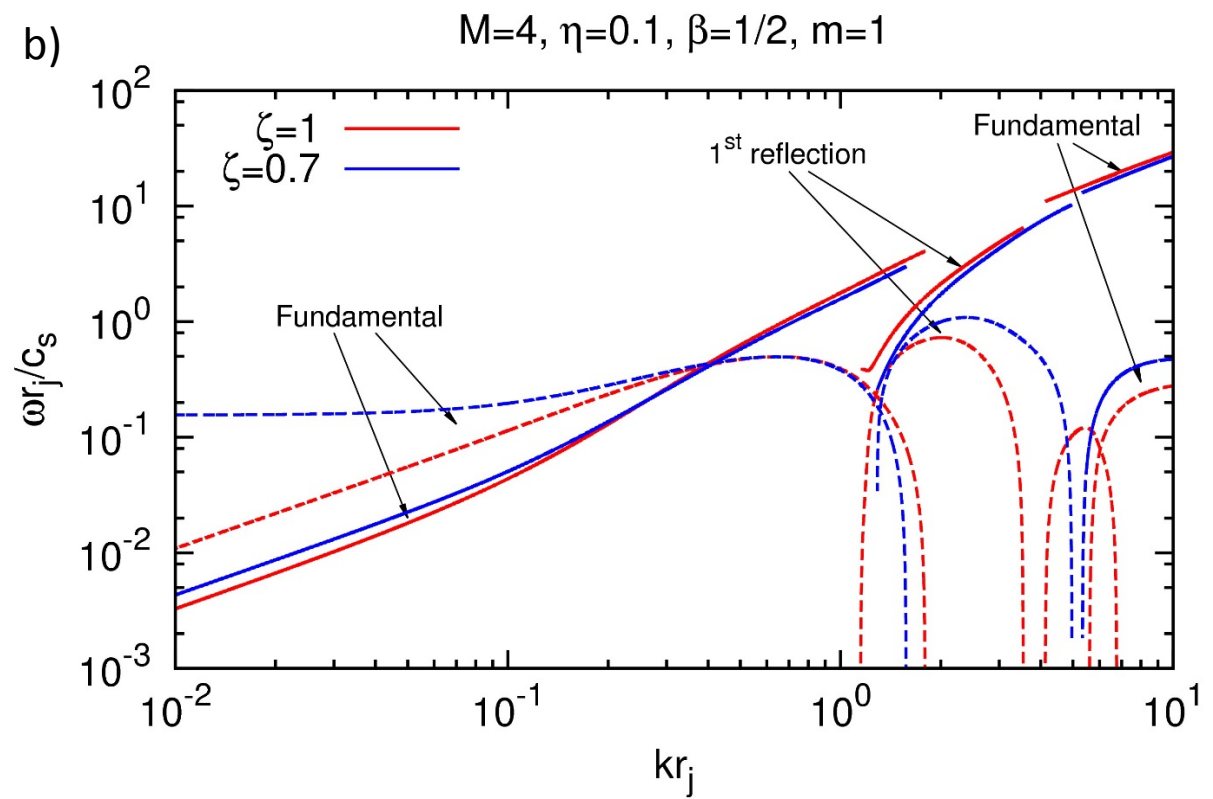

*Fig. 8 inter-compares two magnetized jets with M=4, η=0.1 and β=1/2. The ξ=1.0 case, shown in red, is just the current-sheet free jet that we have studied before in Fig. 5. The ξ=0.7 case, shown in blue, has a current sheet at its boundary. Fig. 8a shows the stability of the m=0 mode. Fig. 8b shows the stability of the m=1 mode. Both figures show the fundamental and first reflection modes. For the m=1 kink mode, the fundamental mode is much more stable for the current-sheet free jet especially at longer wavelengths.*

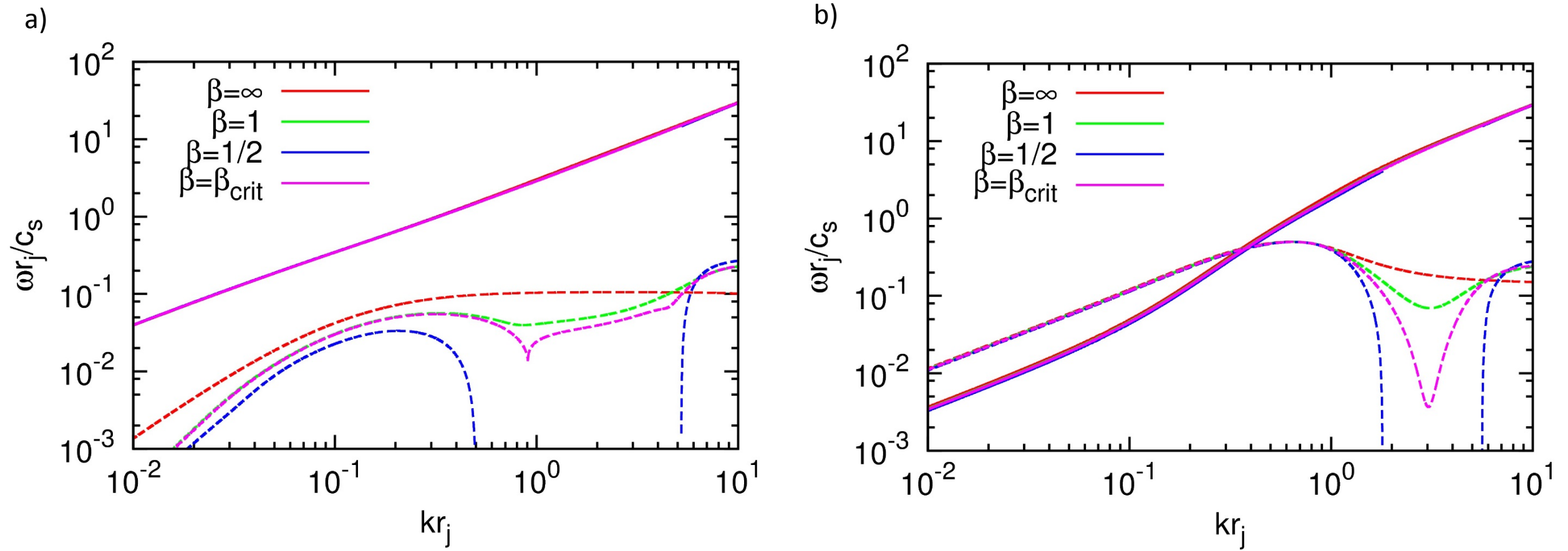

*Fig. 9 we show the fundamental modes for the M=4, η=0.1 jets with β=infinity, β=1, β= β_crit and β=1/2. I.e., the different colors show the enhancement of jet stability as the magnetic field is increased. In all cases the current-sheet free magnetic configuration was used. Fig 9a shows the m=0 fundamental mode while Fig. 9b shows the m=1 fundamental mode. β_crit is the β values when the separation of modes starts to occur and their values are 0.95 (m=0) and 0.76 (m=1) The real parts of the angular frequency, when they are plotted, overlie each other in both the figures.*

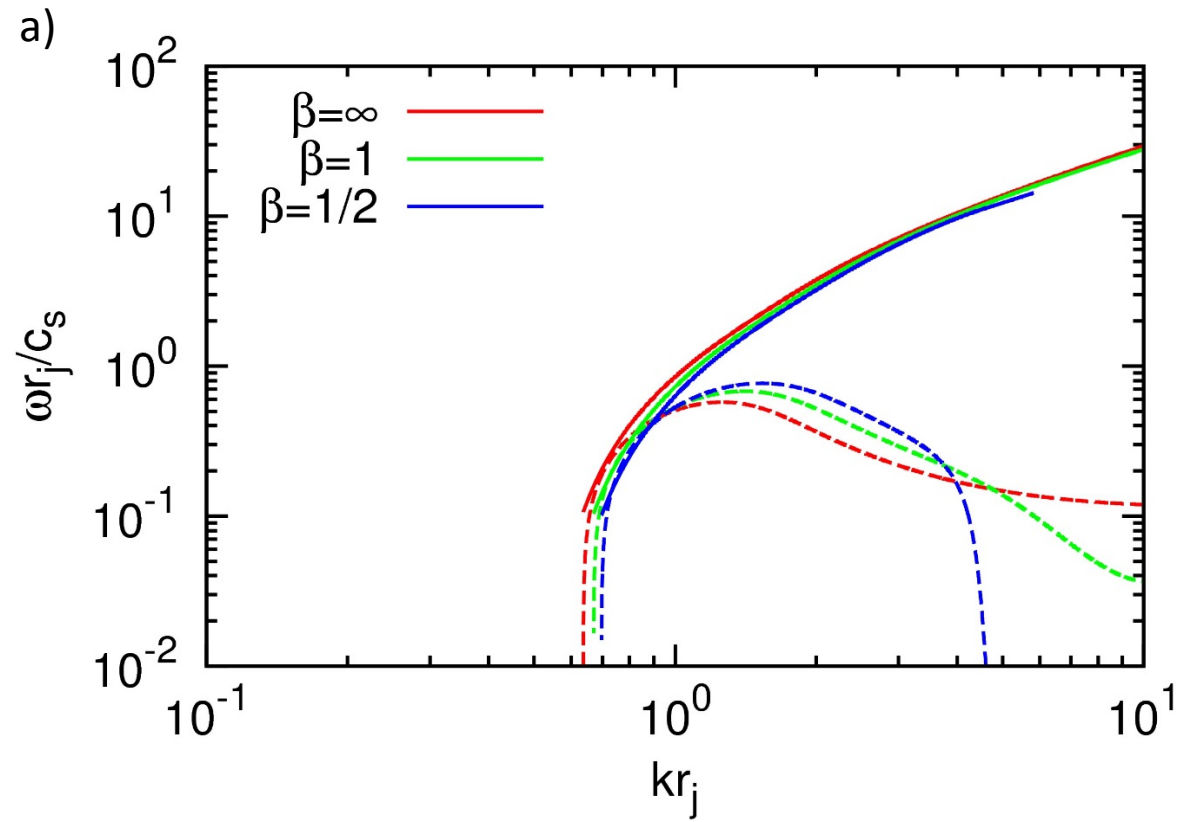 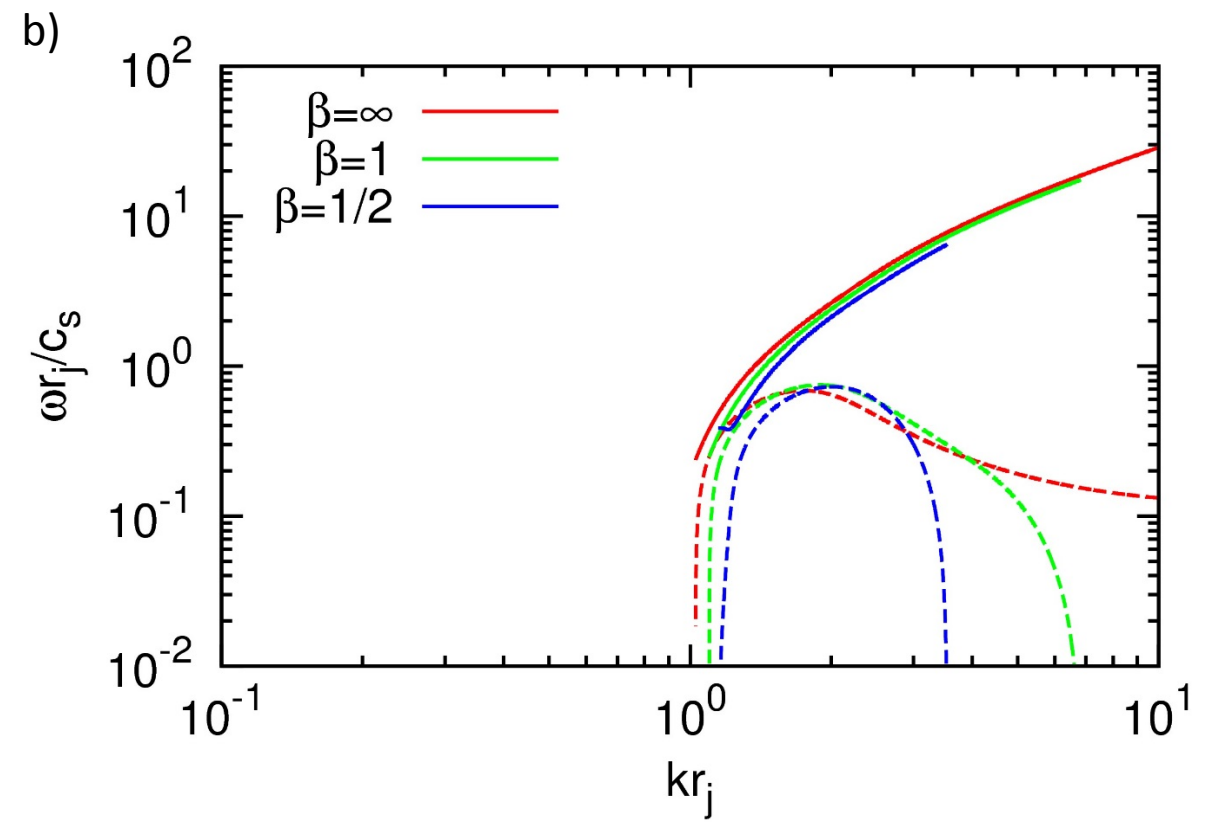

Fig. 10 we show the first reflection modes for the M=4, η=0.1 jets with ß=infinity, ß=1 and ß=1/2. I.e., the different colors show the enhancement of jet stability in the reflection modes as the magnetic field is increased. In all cases the current-sheet free magnetic configuration was used. Fig 10a shows the m=0 first reflection mode while Fig. 10b shows the m=1 first reflection mode.

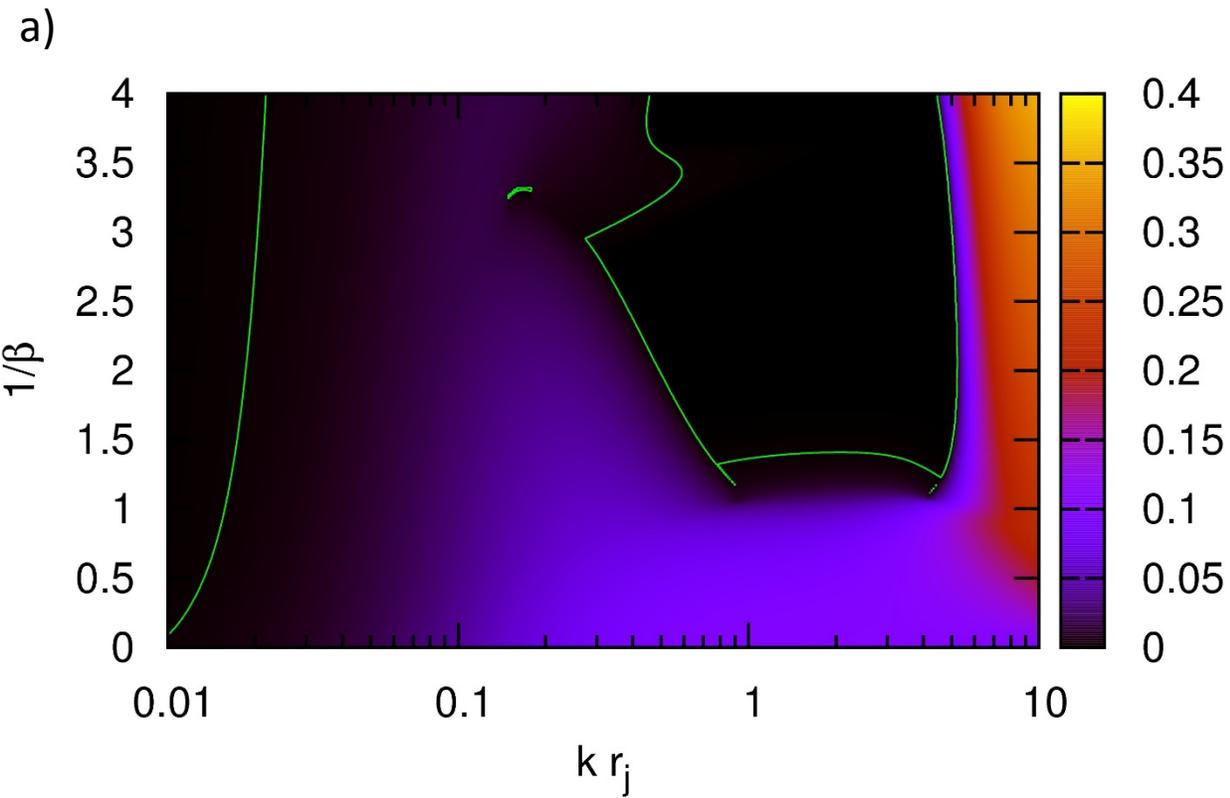 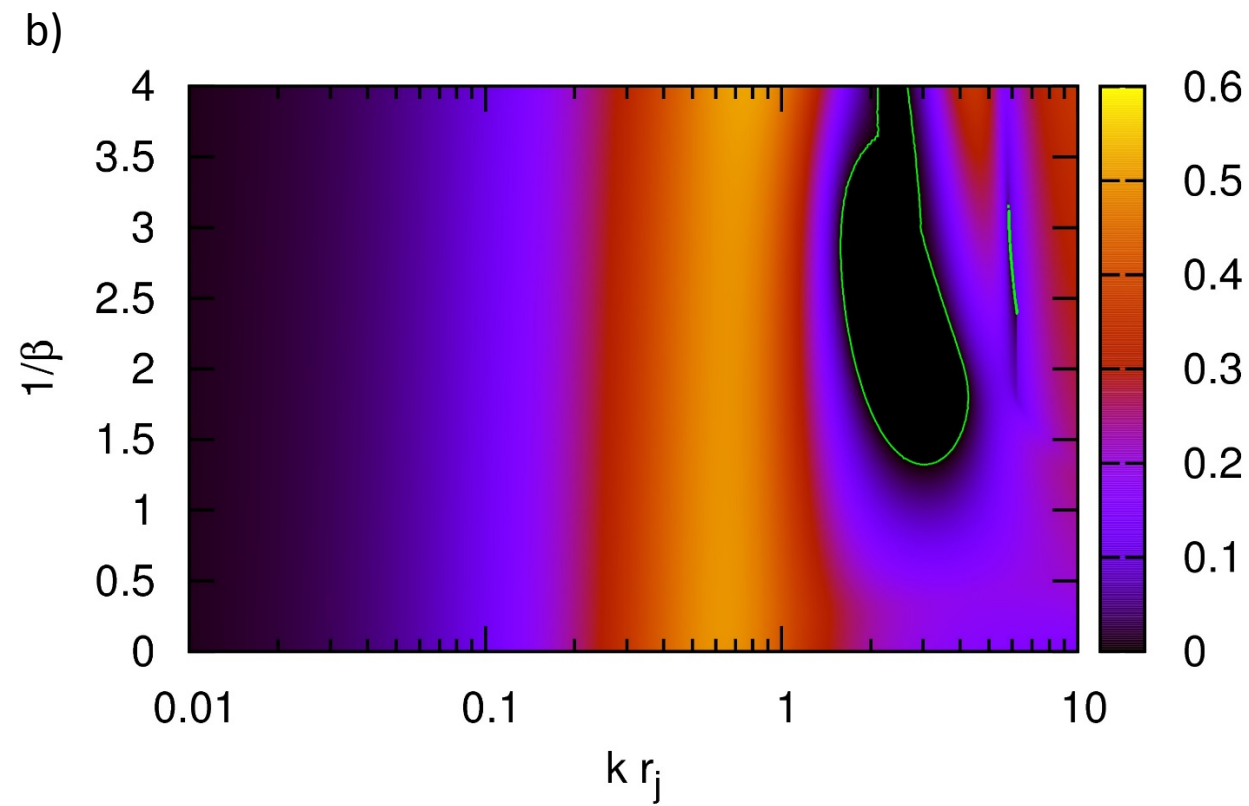

Fig. 11 The imaginary part of the fundamental modes is color coded and shown in Fig. 11. Consequently, Fig. 11a shows the color coded value of the imaginary part of the angular frequency, i.e. $\omega_I\, r_j\, /\, c_s$ as a function of wavenumber and increasing magnetic field (denoted by $1/\beta$) for the $m=0$ fundamental mode. Fig. 11b shows the same information for the $m=1$ fundamental mode. The green lines in Figs. 11a and 11b identify the boundary of the regions past which $\omega_I\, r_j\, /\, c_s$ drops below a value of $10^{-3}$.

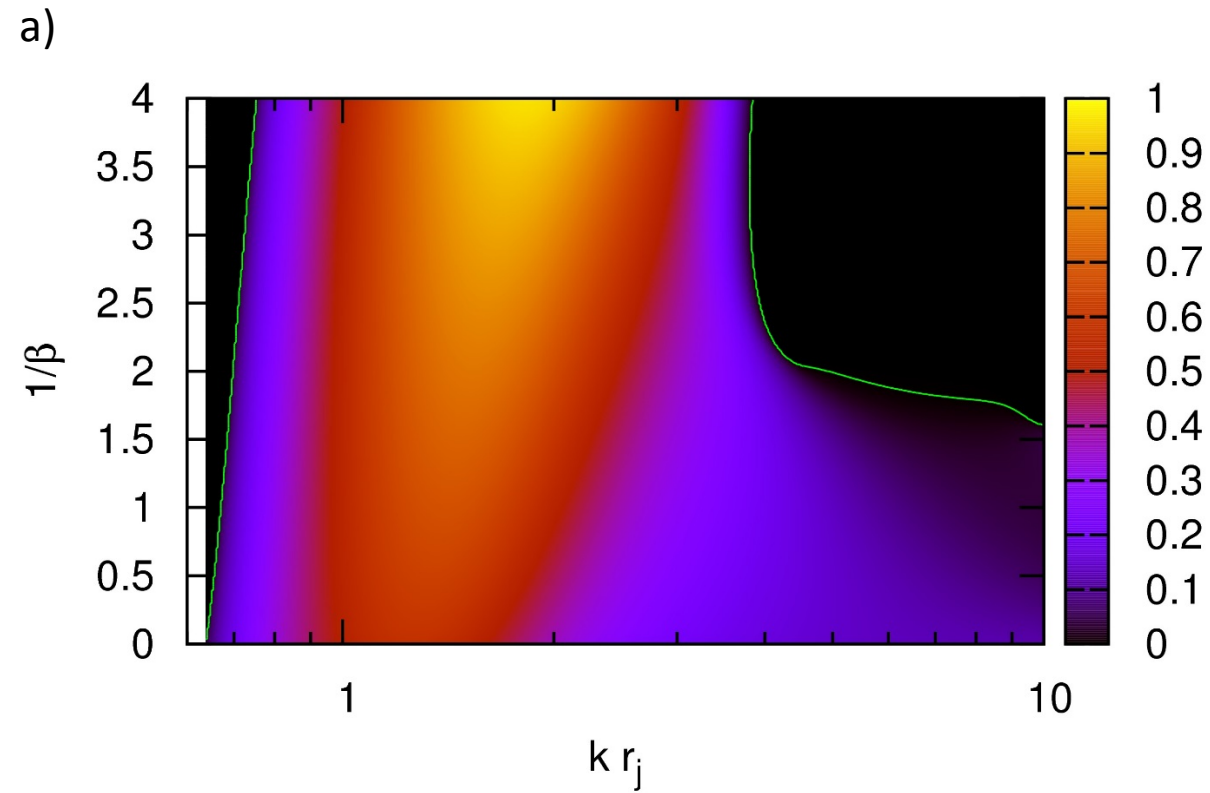 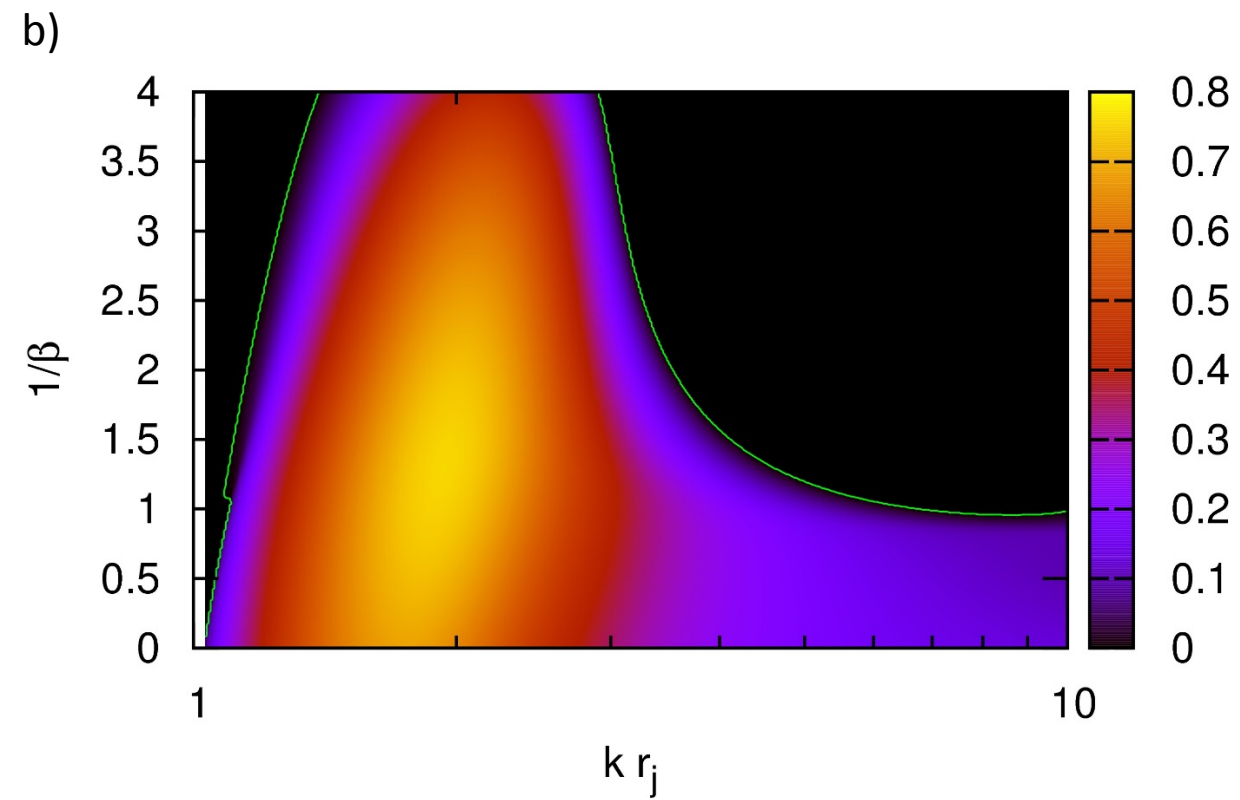

Fig. 12 The same exercise as in Fig. 11 can now be repeated for the first reflection mode. Fig. 12a is analogous to Fig. 11a with the exception that it shows the color coded imaginary part of the angular frequency for the $m=0$ first reflection mode. Similarly, Fig. 12b is analogous to Fig. 11b and shows the color coded imaginary part of the angular frequency for the $m=1$ first reflection mode. The green lines in Figs. 12a and 12b identify the boundary of the regions past which $\omega_I r_j / c_s$ drops below a value of $10^{-3}$.

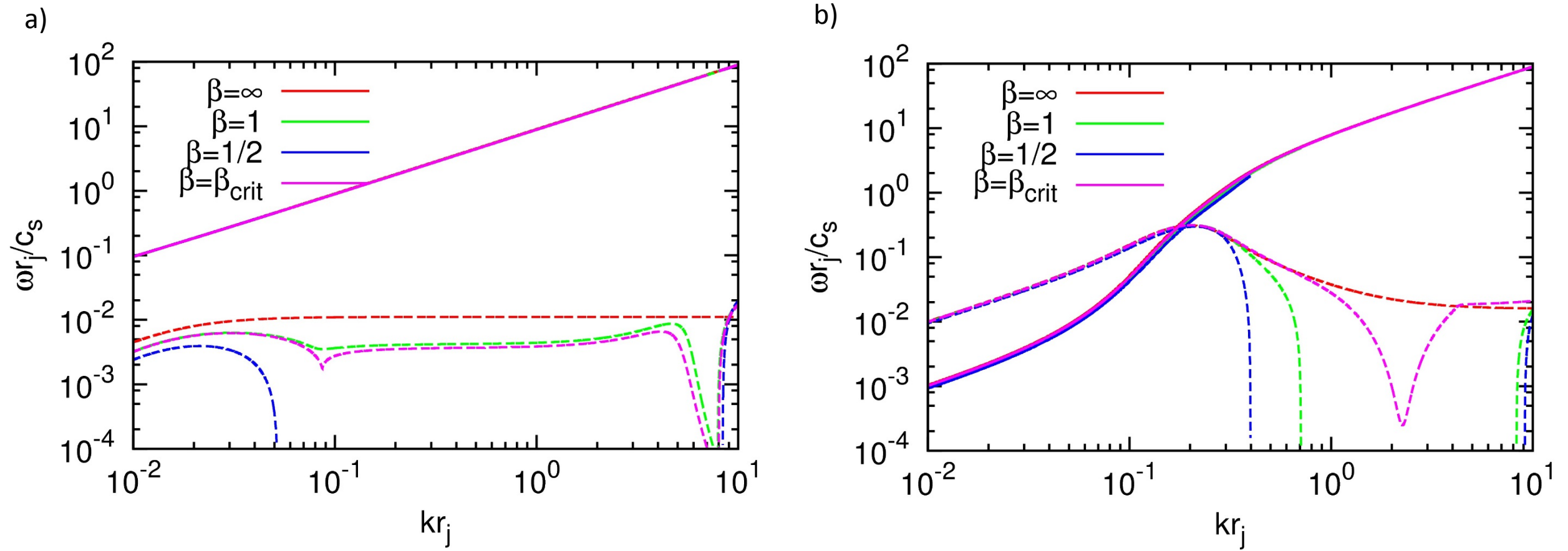

Fig. 13 we study a current-sheet free jet with $M=10$, $\eta=0.01$ and a range of plasma betas. The results of stability analysis for the fundamental $m=0$ and $m=1$ modes are shown in Figs. 13a and 13b. Increasing the strength of the current-sheet free magnetic field dramatically improves the stability properties of light jets. Please compare this figure to Fig. 9.

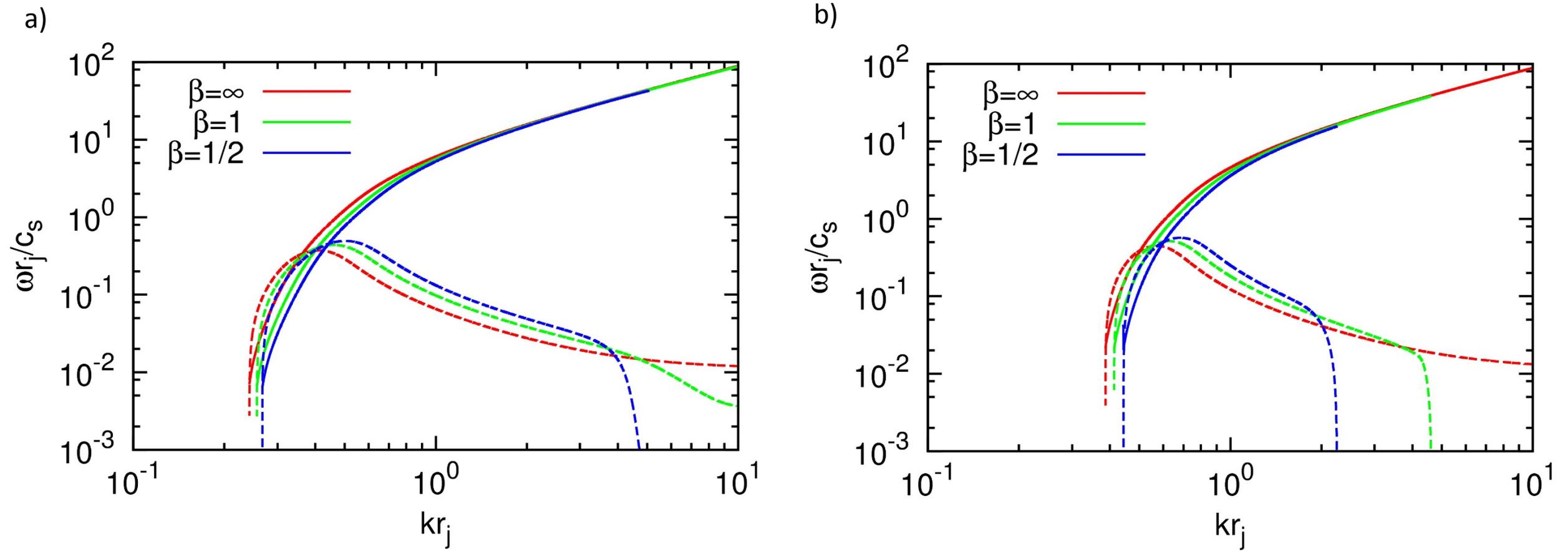

Fig. 14 we study a current-sheet free jet with M=10, η=0.01 and a range of plasma betas. The results of stability analysis for the 1st reflection m=0 and m=1 modes are shown in Figs. 14a and 14b. Increasing the strength of the current-sheet free magnetic field dramatically improves the stability properties of light jets. Please compare this figure to Fig. 10.

a) 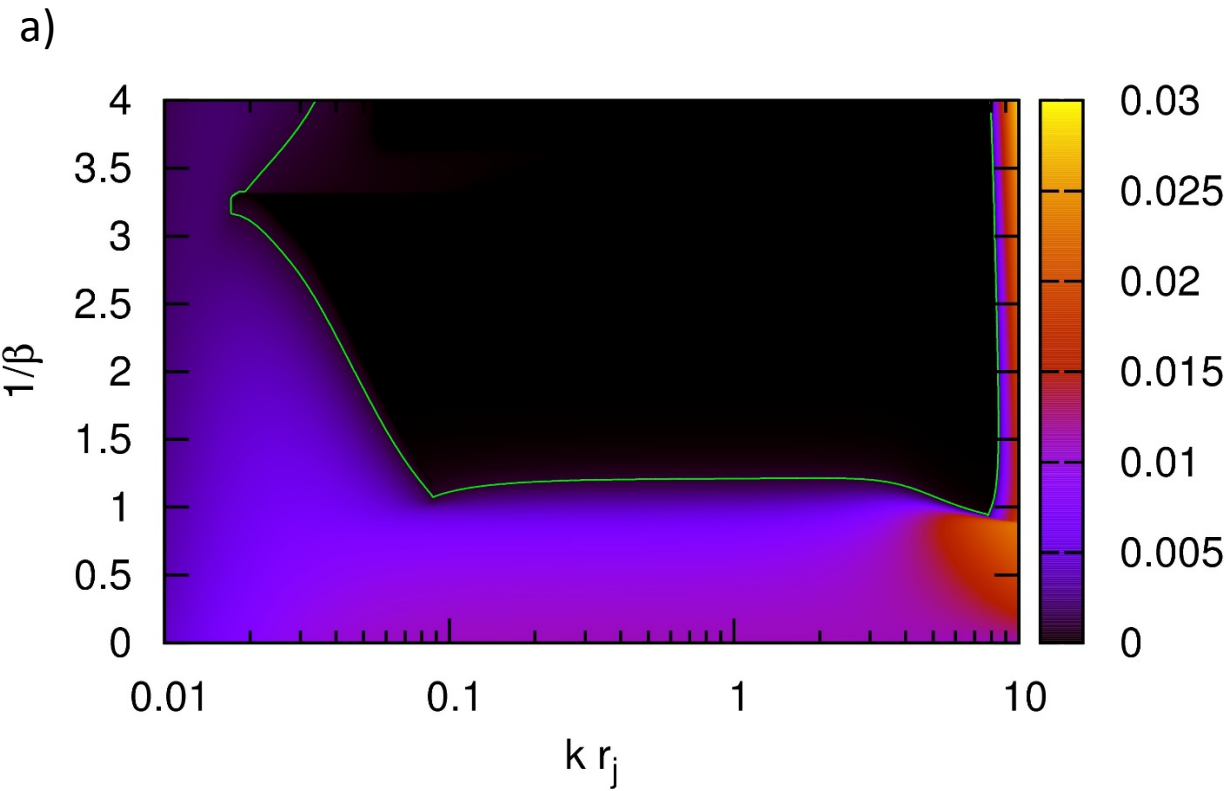

b) 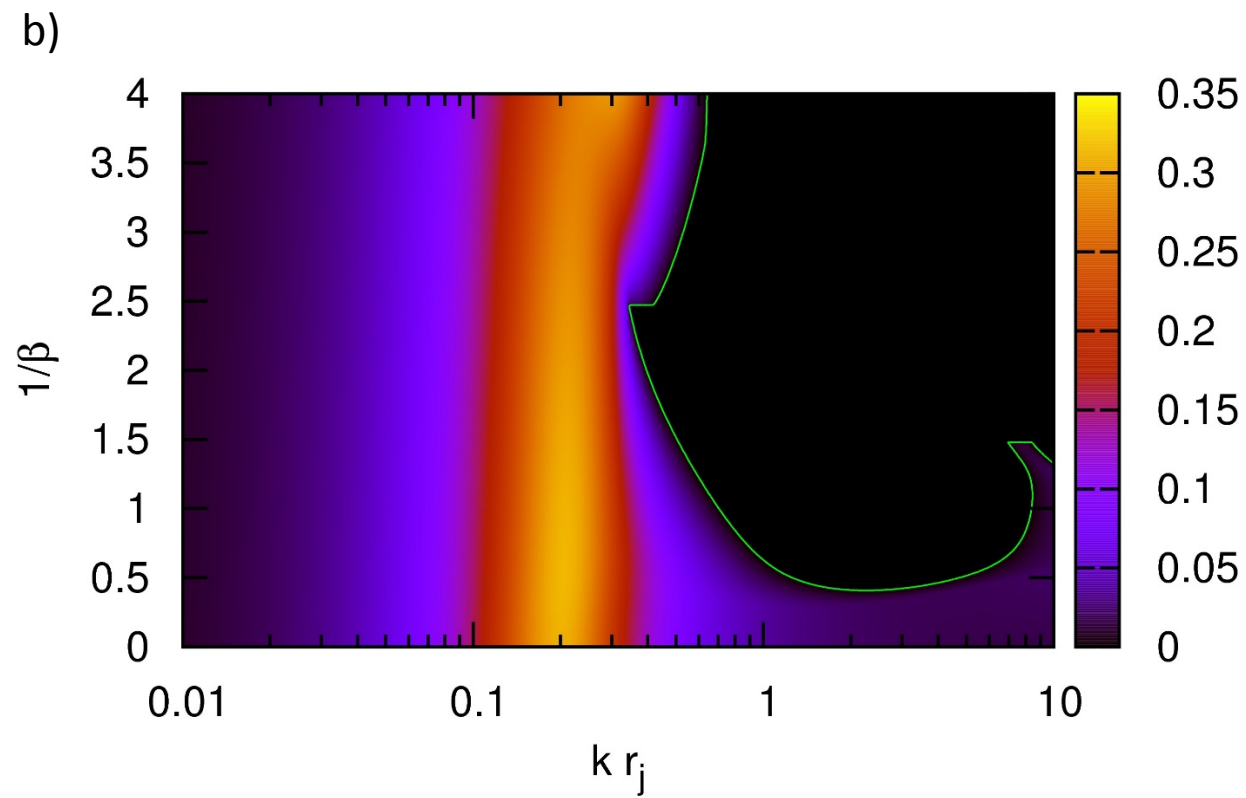

The imaginary part of the fundamental modes is color coded and shown in Fig. 15. Consequently, Fig. 15a shows the color coded value of the imaginary part of the angular frequency, i.e. $\omega_I \, r_j / c_s$ as a function of wavenumber and increasing magnetic field (denoted by $1/\beta$) for the $m=0$ fundamental mode. Fig. 15b shows the same information for the $m=1$ fundamental mode. The green lines in Figs. 15a and 15b identify the boundary of the regions past which $\omega_I \, r_j / c_s$ drops below a value of $10^{-3}$.

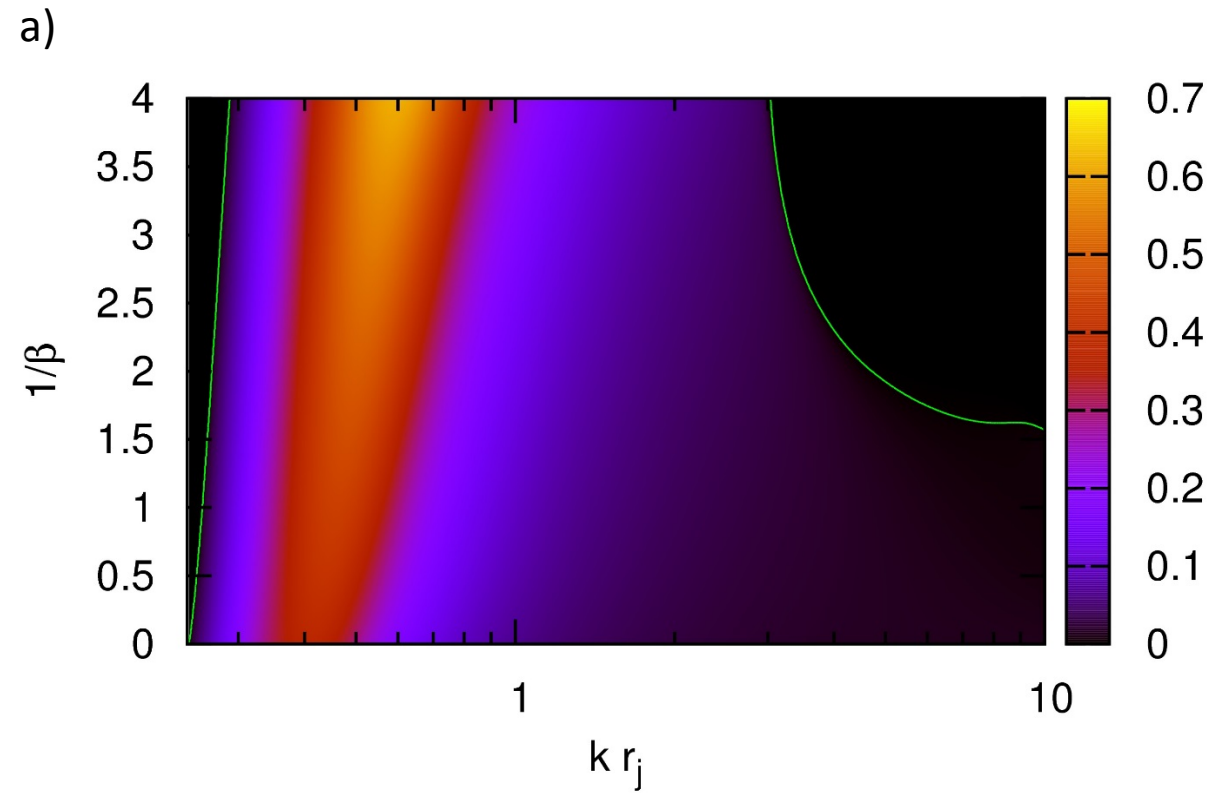 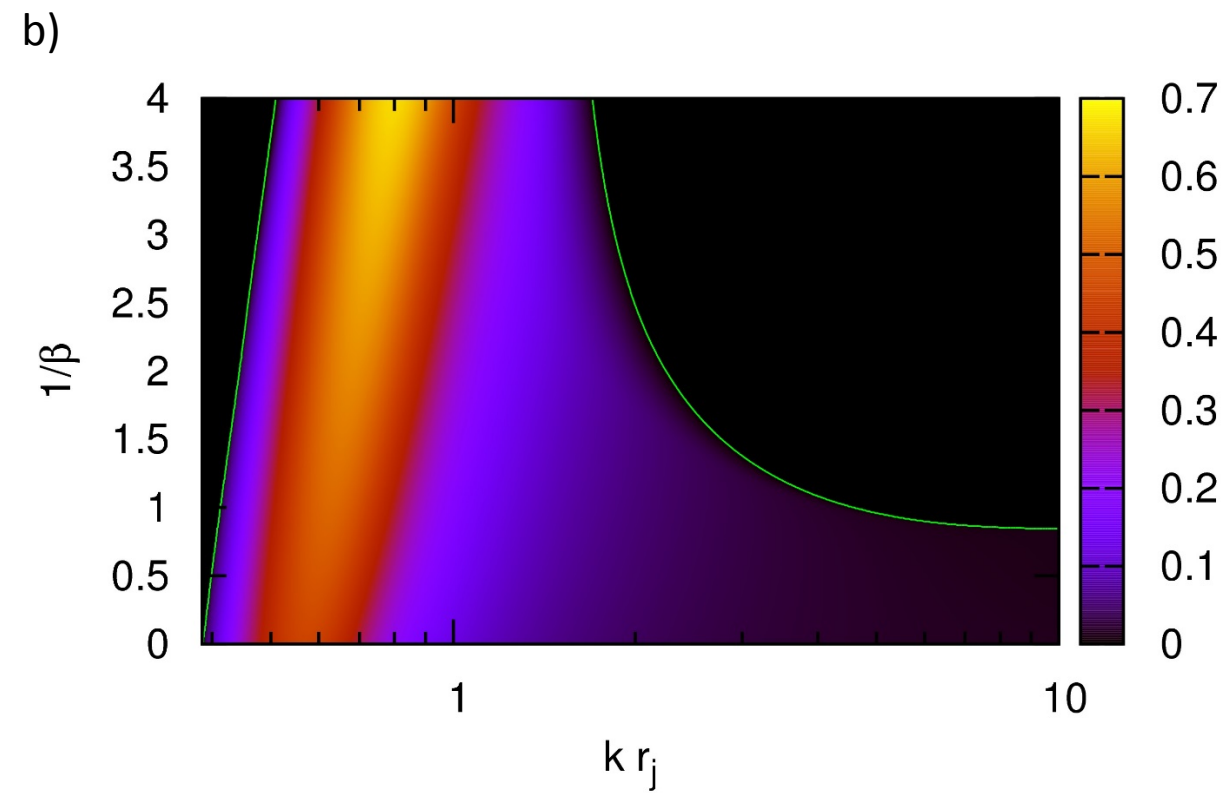

*Fig. 16a is analogous to Fig. 15a with the exception that it shows the color coded imaginary part of the angular frequency for the m=0 first reflection mode. Similarly, Fig. 16b is analogous to Fig. 15b and shows the color coded imaginary part of the angular frequency for the m=1 first reflection mode. The green lines in Figs. 16a and 16b identify the boundary of the regions past which $h\,\omega_I\,r_j\,/\,c_s$ drops below a value of $10^{-3}$.*

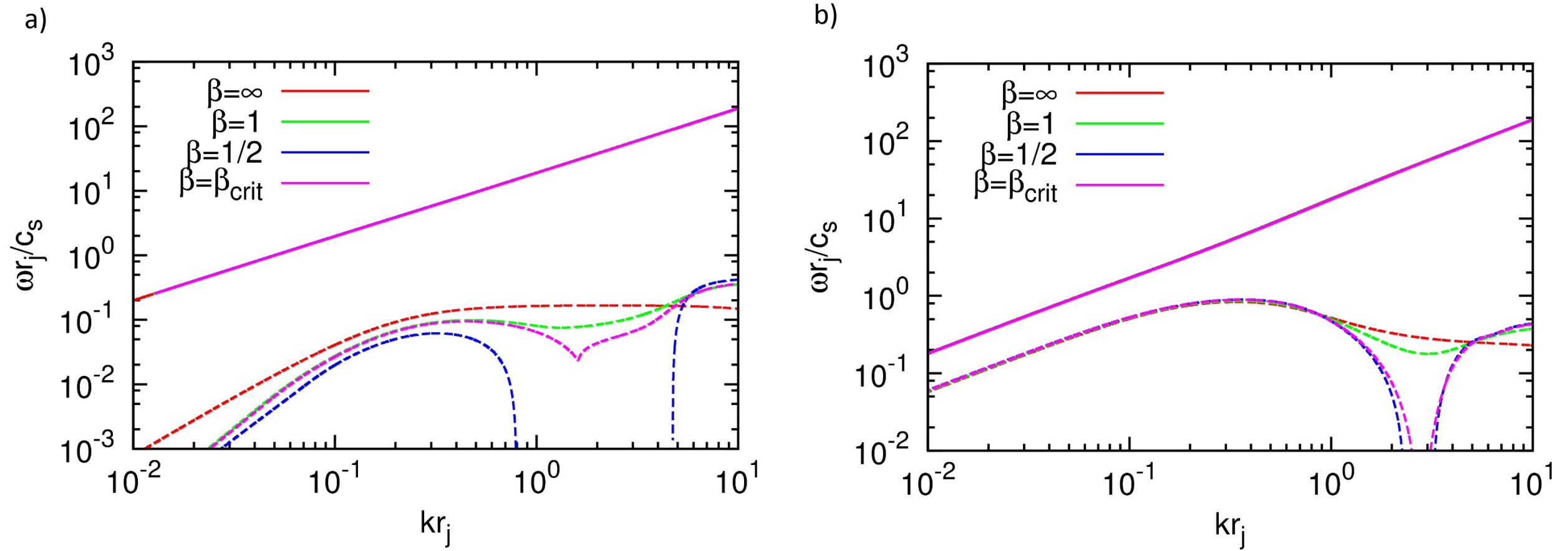

*Fig. 17 The results of stability analysis of a current-sheet free jet with M=20, η=10 and a range of plasma betas for the fundamental m=0 and m=1 modes are shown in Figs. 17a and 17b. By comparing this figure to Figs. 9 and 13 we see that the stability of the heavy jet is not improved as much as the stability of the light jets when the current-sheet free magnetic field is increased.*

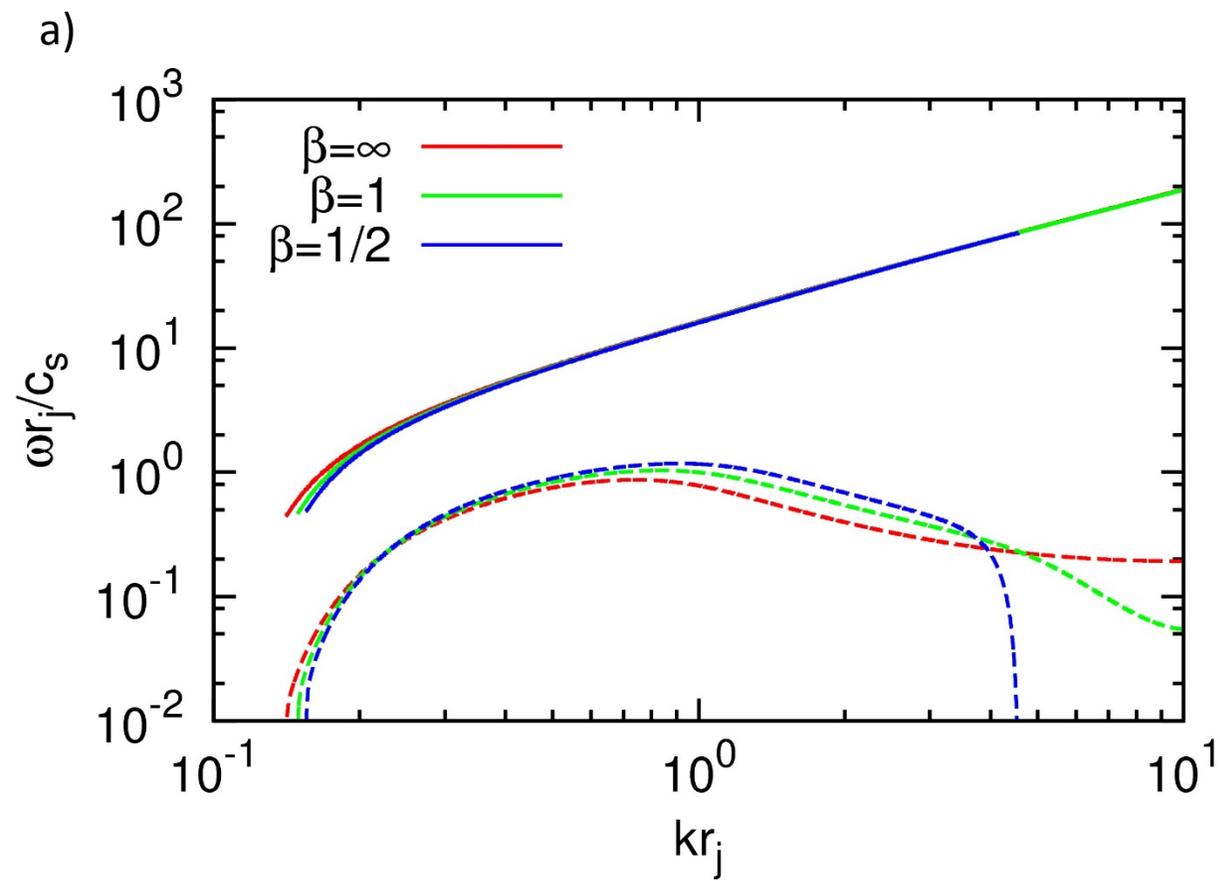 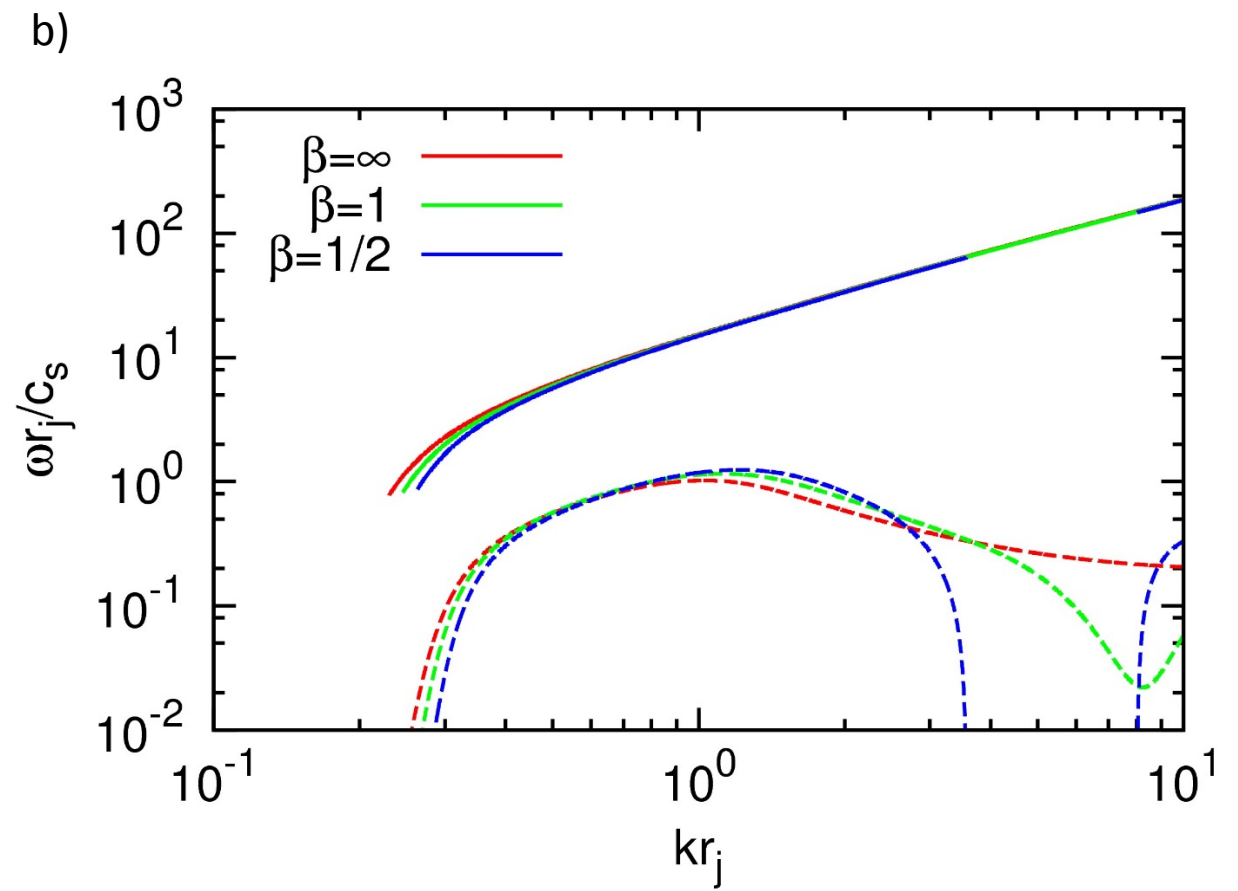

*Fig. 18 The results of stability analysis of a current-sheet free jet with M=20, η=10 and a range of plasma betas for the 1st reflection m=0 and m=1 modes are shown in Figs. 18a and 18b.*

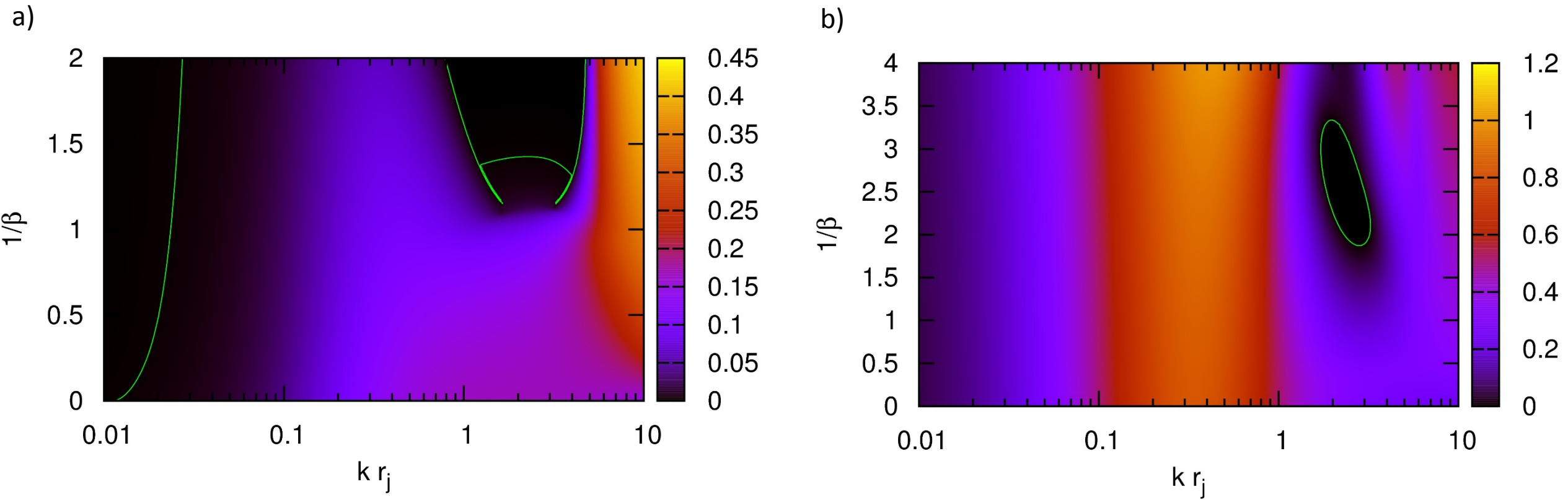

Fig. 19 The imaginary part of the fundamental modes is then color coded and shown in Fig. 19. Consequently, Fig. 19a shows the color coded value of the imaginary part of the angular frequency, i.e. $\omega_I r_j / c_s$ as a function of wavenumber and increasing magnetic field (denoted by $1/\beta$) for the $m=0$ fundamental mode. Fig. 19b shows the same information for the $m=1$ fundamental mode. The green lines in Figs. 19a and 19b identify the boundary of the regions past which $\omega_I r_j / c_s$ drops below a value of $10^{-3}$.

a) 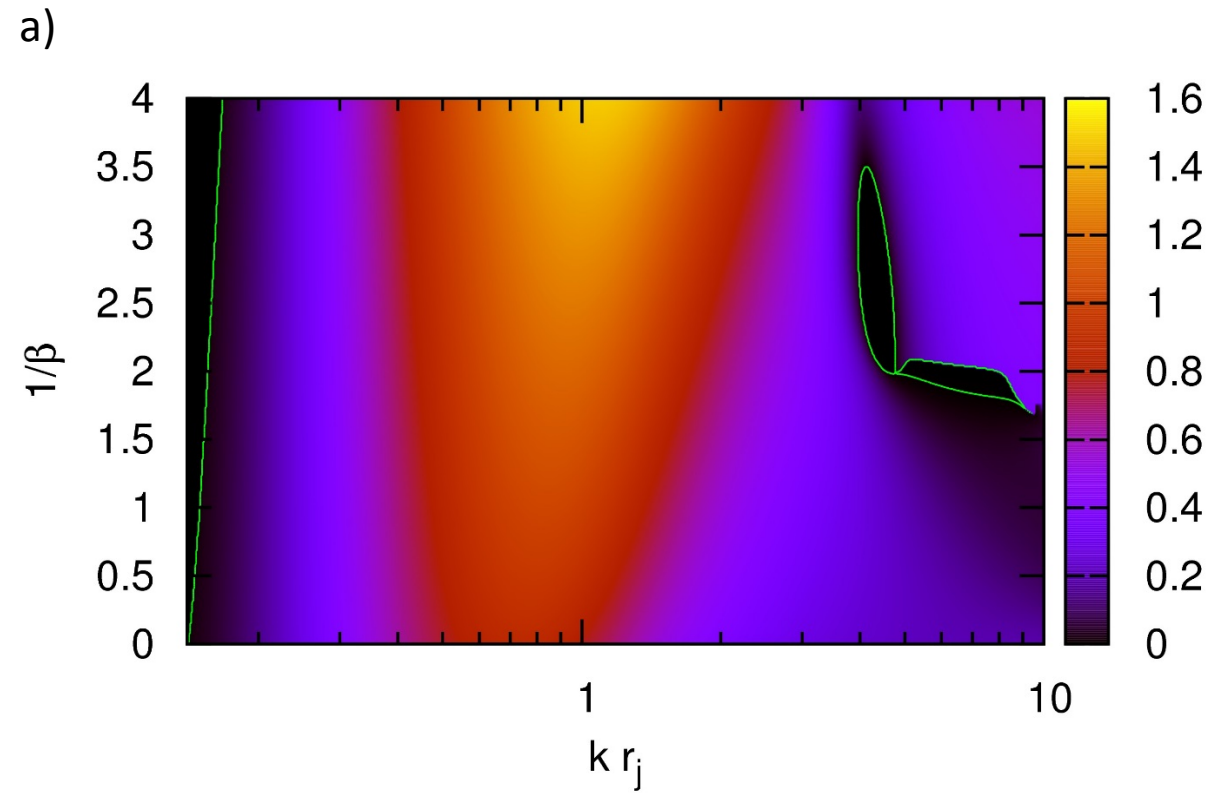 b) 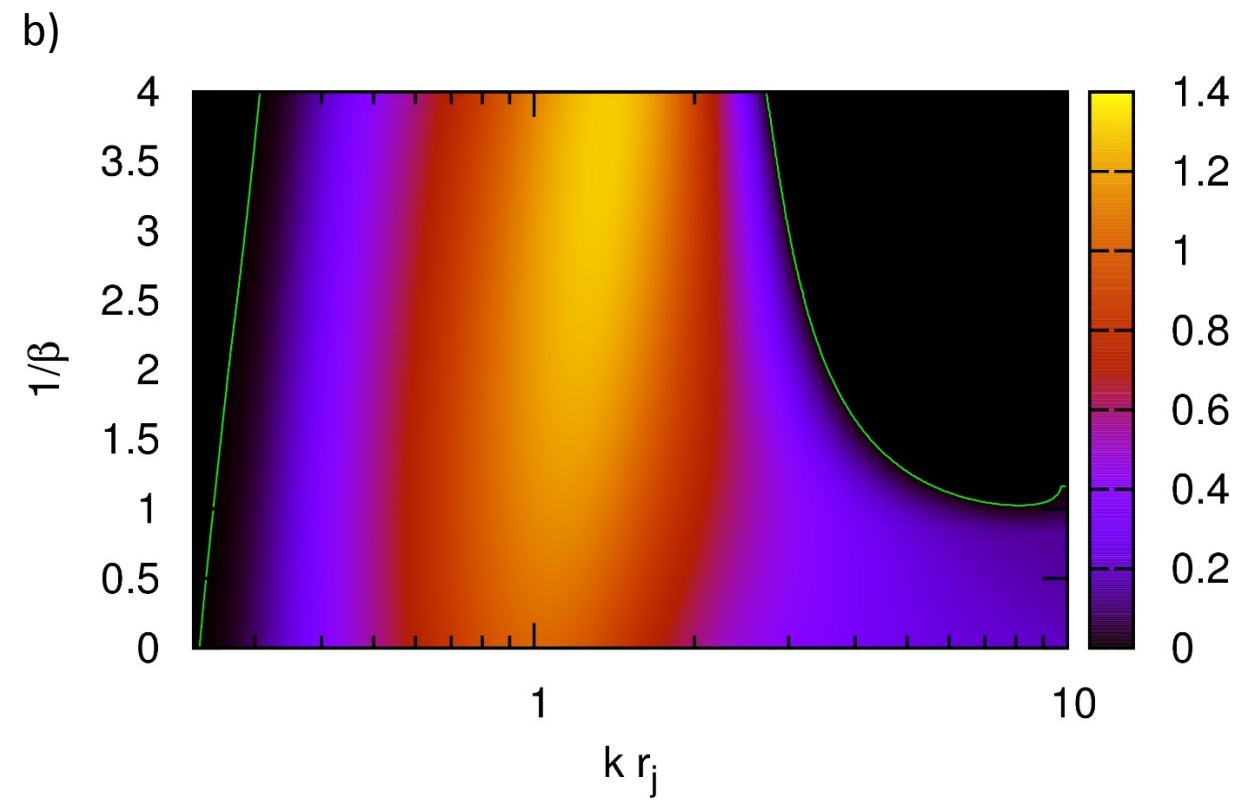

Fig. 20 The same exercise can now be repeated for the first reflection mode. Fig. 20a is analogous to Fig. 19a with the exception that it shows the color coded imaginary part of the angular frequency for the m=0 first reflection mode. Similarly, Fig. 20b is analogous to Fig. 19b and shows the color coded imaginary part of the angular frequency for the m=1 first reflection mode. The green dashed lines in Figs. 20a and 20b identify the boundary of the regions past which $\omega_I r_j / c_s$ drops below a value of $10^{-3}$.